\documentclass[
 aip,
 pop,
 amsmath,amssymb,
 reprint,
 onecolumn
]{revtex4-1}

\usepackage{graphicx}
\usepackage{dcolumn}
\usepackage{bm}
\usepackage{amsthm}
\usepackage[utf8]{inputenc}
\usepackage[T1]{fontenc}
\usepackage{mathptmx}
\usepackage{etoolbox}

\newenvironment{subfigure}[1]{\begin{minipage}{#1}}{\end{minipage}}

\makeatletter
\def\@email#1#2{%
 \endgroup
 \patchcmd{\titleblock@produce}
  {\frontmatter@RRAPformat}
  {\frontmatter@RRAPformat{\produce@RRAP{*#1\href{mailto:#2}{#2}}}\frontmatter@RRAPformat}
  {}{}%
}
\makeatother

\begin{document}

\title{Linear Gyrokinetic Simulations of Micro-tearing Mode: Local versus Global}

\author{Yifei Liu}
\affiliation{Southwestern Institute of Physics, Chengdu 610041, China}%
\author{Haotian Chen}%
 \email{chenhaotian@swip.ac.cn}
\affiliation{Southwestern Institute of Physics, Chengdu 610041, China}%
\author{Yao Yao}
\affiliation{Southwestern Institute of Physics, Chengdu 610041, China}%
\author{Zhengji Li}
\affiliation{Southwestern Institute of Physics, Chengdu 610041, China}%
\author{Jiquan Li}
\affiliation{Southwestern Institute of Physics, Chengdu 610041, China}%
\author{Wei Chen}
\affiliation{Southwestern Institute of Physics, Chengdu 610041, China}%

\date{July 17, 2026}

\begin{abstract}
A systematic comparison of local and global linear gyrokinetic simulations of micro-tearing modes (MTMs) is performed using the GENE code.
The analysis spans diverse plasma parameters, including the core regions with normal and weak magnetic shear, as well as the pedestal region with the strong plasma non-uniformity.
The global simulations reveal a distinct MTM type characterized by a `parity mixing' mode structure, which can be significantly destabilized by trapped electrons.
% Unlike the standard tearing parity MTMs, where the trapped electron response is weakened by bounce dynamics across the odd parity mode structure, this parity mixing mode corresponds to local simulations with finite ballooning angles.
Moreover, in contrast to electrostatic drift wave instabilities, the current layer width ($ \Delta _c $) is identified as the crucial factor determining the importance of global effects.
The MTM in the core region exhibits the slab-like feature with narrow $ \Delta _c $, leading to high consistency between local and global results.
However, in the pedestal region, the steep pressure gradient broadens $\Delta_c$, driving quantitative deviations when $\Delta_c$ becomes comparable to the plasma pressure gradient scale length.
For high-$n$ MTMs, $ \Delta _c $ can exceed the distance between adjacent mode rational surfaces. The resulted overlapping of current layers enhances the toroidal mode coupling effect, accounting for the substantial discrepancies observed between local and global simulations.
% The results indicate that the slab-like feature of MTMs, characterized by weakly coupled harmonics, is the primary reason for their insensitivity to global profile variations and the resultant agreement between local and global results in core plasmas.
% However, significant deviations between local and global results emerge in pedestal region, where the strong non-uniformity alters the current layer of MTMs.
% Moreover, we report an MTM exhibiting `parity mixing' harmonic structures and identify it as a toroidal mode.
% Its destabilization mechanism arises from trapped electrons at a finite ballooning angle ($\theta_k \neq 0$), which is absent in $\theta_k=0$ local simulations due to the tearing parity.
% Another type of toroidal MTM is found in the pedestal region, where the coupled current layers form radially extended structures. This renders the mode sensitive to the variations of the equilibrium profiles, leading to qualitative discrepancies between local and global results.
% These findings not only delineate the limitations of local MTM analysis but also uncover significant destabilization mechanisms. %, highlighting the importance of global effects and finite $ \theta _k $ local parameters for a predictive understanding of MTM stability.
\end{abstract}

\maketitle

\section{Introduction}

Future fusion reactors will rely mainly on alpha particles for plasma heating, a process that primarily transfers energy to electrons.
This highlights the critical role of electron driven microturbulence in driving anomalous electron transport.
Since high-$ \beta $ plasmas are essential for economic efficiency, electromagnetic micro-instabilities such as micro-tearing modes (MTM) have attracted growing interest.
The MTM is a tearing parity mode characterized by even symmetry of poloidal harmonics of perturbed parallel magnetic potential, $ \delta A_{\|,m} $, with respect to its mode rational surface (MRS).
In the ballooning mode representation, this parity manifests as the even symmetry of $\delta A_{\|}$ along the extended poloidal angle.
The tearing parity of the MTM gives rise to the formation of microscopic magnetic islands.
However, unlike macroscopic tearing modes, the MTM is driven by the electron temperature gradient rather than the current gradient.
Consequently, it induces significant electron thermal transport \cite{Rechester1978} without causing plasma disruption.
The MTM is typically characterized by fine radial structures of the perturbed electrostatic potential ($\delta \phi$) and the current layers.
These structures correspond to the widely extended mode structures observed in the extended poloidal angle space of local simulations, which is extensively utilized for MTM analyses\cite{Applegate2007,Wong2007,Guttenfelder2011,Doerk2011,Chowdhury2016,Maeyama2017,Jian2019,Pueschel2020,Giacomin2023}.
Notably, early studies in the literature \cite{Doerk2012} suggest that, in contrast to the ion temperature gradient mode (ITG) \cite{Candy2004, McMillan2010, Villard2010}, MTM is generally insensitive to radial profile variations.
However, recent works focusing on the micro-instabilities in the pedestal region report a significant discrepancy between local and global results \cite{Chen2018,Hatch2021,Larakers2021,Hassan2021}.
To clarify the compatibility between properties of MTM and the local simulations across different plasma regimes,
a systematic verification is required.

In this paper, a comparison between local and global simulations of MTMs is conducted for both core and pedestal plasma regions.
% In contrast to electrostatic drift wave instabilities\cite{Chen2018}, the significance of global effect of MTMs is fundamentally controlled by the current layer width ($\Delta_c$).
% we reveal that the divergence between global and local MTM simulations is intrinsically determined by the current layer width ($\Delta_c$).
The current layer width ($\Delta_c$) of MTMs is identified as the crucial parameter for the divergence between global and local MTM simulations.
This effect is particularly pronounced in the pedestal region, where the steep pressure gradient broadens $ \Delta _c $ to the scale length of the plasma pressure gradient, leading to quantitative deviations between local and global results.
Moreover, the overlapping of current layers from adjacent MRSs are observed in high toroidal mode number (high-$n$) MTMs when the separation of MRSs ($ \Delta _m $) becomes narrower than $ \Delta _c $.
The overlapping introduces toroidal mode coupling and complex dependencies on equilibrium profiles, leading to substantial discrepancies between local and global simulation results.
For core plasmas, the slab-like feature and narrow $ \Delta _c $ of MTMs are identified as the principal cause for observed insensitivity of the mode to the profile variations.
However, a distinct MTM type characterized by a ‘parity mixing’ structure, which is significantly destabilized by trapped electrons, is observed in global simulations.
This parity mixing structure corresponds to local simulations with finite ballooning angles ($ \theta _k $)\cite{Chen2018a}.
In contrast, the response of trapped electron to the MTM with zero $ \theta _k $ is weakened by the bounce averaging over the odd parity mode structure.
The mode is identified as a toroidal mode, thereby increasing its sensitivity to profile variations.

This paper is structured as follows.
A detailed description of the local and global simulation models utilized in this research is provided in Section \ref{sec:Methodology}.
Sections \ref{sec:Normal Gradient} and \ref{sec:Strong Gradient} present the simulations of the MTMs in core and pedestal regions, respectively.
Finally, a summary is provided in Section \ref{sec:Summary}.

\section{Methodology}\label{sec:Methodology}
 
The global gyrokinetic simulation solves gyrokinetic Vlasov-Maxwell equations \cite{Frieman1982,Brizard2007,Chen2024} in a given radial region with profile variations,
its boundary is artificial yet physically appropriate for the concerned physics.
For example, in the ITG simulations at core plasmas, where the mode is excited at center of simulation region,
the fixed boundary condition (B.C.) is usually adapted in many gyrokinetic simulation codes \cite{Lin1998,Chen2007,Wang2018}.
However, the global simulation requires a massive computational resources in micro-turbulence simulation, especially when kinetic electrons' response is considered.

In contrast, the local simulation, also called the flux-tube simulation \cite{Beer1995,Kotschenreuther1995,Jenko2000,Chen2003,Candy2003}, can reduce computational cost of the simulation effectively. 
It assumes a simulation region along magnetic field line called flux tube that statistical identical to other flux tubes.
For consistency, both the periodic B.C. and local assumption of equilibrium profiles must be satisfied.
Specifically, the simulation region is periodic in the radial and binormal directions,
and the radial width $ L_x $ of the region must be an integer multiple of $ \Delta_m $.
% The local assumption omits equilibrium profile variations of the safety factor $ q $, density $ n $, temperature $ T $ and metric coefficients.
% Instead, the gradients $ \hat{s}=r/q dq/dr $, $ d n/dr $, $ dT/dr $ are fixed as their values at reference points.
The local assumption treats the simulation domain as a narrow neighborhood around a specific reference radial surface. Consequently, both the background equilibrium parameters (e.g., the safety factor $ q $, density $ n $, temperature $ T $ and metric coefficients) and their corresponding first-order radial gradients (e.g., $ \hat{s}=(r/q) dq/dr $, $ d n/dr $, $ dT/dr $) are evaluated at this reference point and treated as constants throughout the domain.
% With these assumptions, the flux-tube simulation is analogous to a crystal simulation, i.e.,
% the flux tube acts as the equivalent of a primitive cell or supercell and the validity of the simulation is guaranteed by Bloch's theorem.

In this work, we employ both local and global simulations using the GENE code \cite{Jenko2000,Goerler2011}, 
which solves the gyrokinetic Vlasov-Maxwellian equations on a fixed grid in $ \{x,y,z,v_\|,\mu \}$ phase space and includes the essential electromagnetic physics and velocity-dependent collision operator for MTM excitation. The code has been tested in numerous local or global electromagnetic simulations in both core and pedestal regions\cite{Doerk2012,Hatch2016,Gorler2016,Hatch2021,Hassan2021,Ajay2023,Curie_2022,Dominski2024}.
The compressional magnetic perturbation $ \delta B_\| $ is neglected in simulations, as it has a negligible impact on MTM physics \cite{Dickinson2013} and presents numerical challenges in solving the global field equations.
% has negligible impact on MTM \cite{Dickinson2013} but is difficult to be solved in field equations of global simulations.
The Dirichlet B.C. is imposed radially in global simulations, and a Krook operator is applied to artificially reduce perturbation levels at the boundaries, which is analogous to the fixed B.C..
% The parameters of the Krook operator are chosen appropriately and have no artificial impact on the physical results.
% This comparison work covers three typical regions of tokamak: core with normal shear ($ R/L_{Te}\simeq 10 $, $ \hat{s}\simeq 1 $), core with weak shear $ \hat{s}\simeq 0.1 $ and pedestal ($ R/L_{Te}\simeq 100 $, $ \hat{s}\simeq 0.1 $).
% We compare the linear results, i.e., the mode frequencies and structures of local and global results in detail, including parameters region of normal or weak magnetic shear and gradients of core or pedestal.
% We present a detailed comparison of the linear results, specifically mode growth rates, frequencies and structures, between local and global simulations. This comparison covers parameter regimes characterized by normal or weak magnetic shear, as well as core and pedestal gradients.
We compare the linear growth rates, frequencies and mode structures obtained from local and global simulations in detail, spanning parameter regimes with normal or weak magnetic shear,
as well as varying degrees of plasma non-uniformity, specifically regarding the magnitude of profile variations in the safety factor, density, temperature, and metric coefficients.

% both the consistence and inconsistence result are researched in detail and a few potential physical mechanisms are firstly reported.

\section{MTM simulations in core plasmas}
\label{sec:Normal Gradient}

\begin{figure}[htbp]
  \begin{center}
    \includegraphics[width=0.315\textwidth]{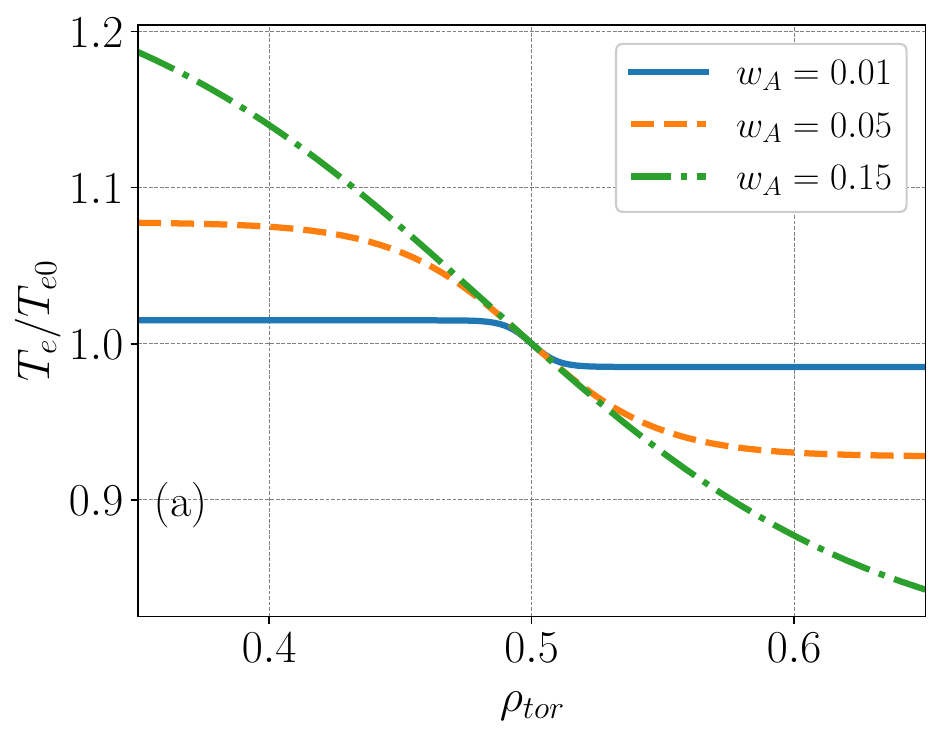}
    \includegraphics[width=0.3\textwidth]{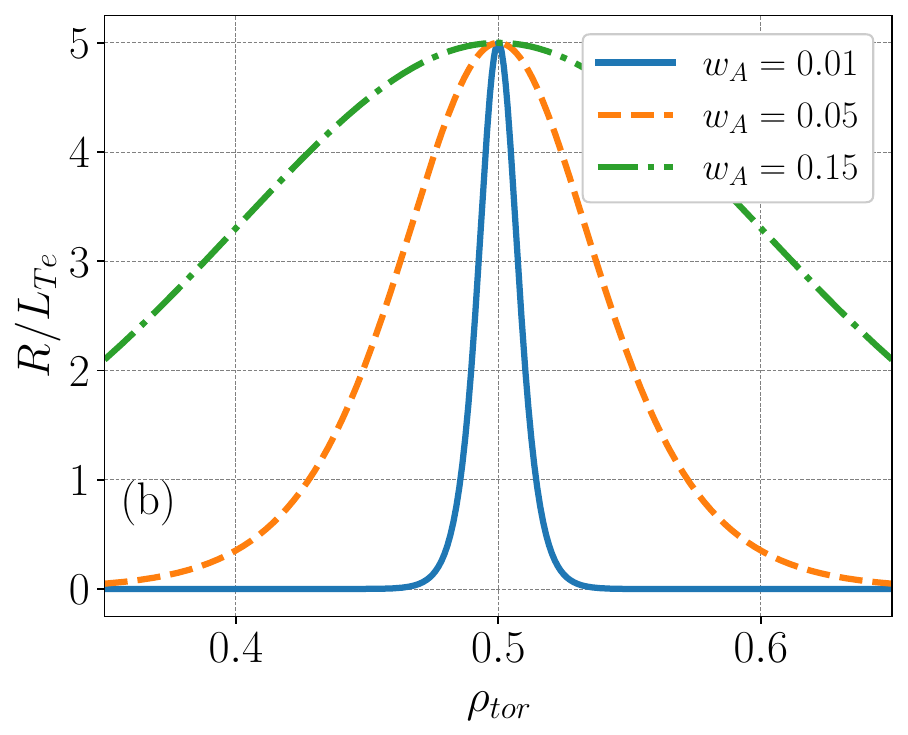}
  \end{center}
  \caption{Profiles of (a) normalized electron temperature $ T_e/T_{e0} $ and (b) the normalized temperature gradient $ R/L_{Te} $ for global simulations of core plasmas. Here $ \rho _{tor} $ is the normalized toroidal magnetic flux under square root, which equals $ x/a $ in circular geometry. Different curves denote varying values of $ w_A $.}
  \label{fig:core_profile}
\end{figure}

For simplicity and thereby clarity, we adopt a concentric circular geometry and neglect the shaping effect, $ \alpha  $ effect and Shafranov shift in simulations. These simplifications are relevant to core plasmas.
% We find two kinds of MTMs in this section, the slab-like MTM and the trapped electrons destabilized MTM.
% Detailed parameter settings will be elaborated on in their respective subsections

% \subsection{Slab-like MTM}
% \label{sub:slabmtm}
For the normal magnetic shear case, parameters are similar to those in Doerk and Ajay's works \cite{Doerk2011,Ajay2023}.
The profiles of density and temperature in global simulations are set by the form of
$ A(x)=A_0\exp \left\{ - \kappa _A w_A a/R \tanh \left[ (x-x_0)/(w_A a) \right] \right\} $, where $A \in \{n, T_e, T_i\}$ and $ T_{e0}=T_{i0} $, $ a $ and $ R $ denote the minor and major radii, respectively, $ w_A $ control the width of the driving gradient region, as shown in Fig. \ref{fig:core_profile}.
$ x_0=0.5a $ serves as the midpoint for global simulations and the reference point for local simulations.
% and $ x_0=0.5a $ is the reference point of local simulations and mid-point of global simulations.
Here, $ q_0=3 $, $ \hat{s}=1 $, $ R/L_{n}= \kappa_n = 1 $, $ R/L_{Ti}=\kappa _{Ti}=0 $, $ R/L_{Te}= \kappa _{Te}=5 $,
$ \beta _{e}=0.8 \% $, the electron-ion collision frequency $ \nu _{ei}=0.86 c_{s0}/R $, $ c_{s0}= \sqrt{T_0/m_H} $, the mass ratio of electrons and ions $ m_e/m_i = m_e/m_H = 1/1836 $, the inverse aspect ratio $ \epsilon = x_0/R=0.15 $.
The $ q $ profile is set as $ q=1.5+6 x^2 $, the radial simulation domain width $ L_x=0.3a $, $ \rho_*= \rho_{s0}/a= 0.004 $, $ \rho _{s0}=m_H c_{s0} /eB_0 $, $ B_0 $ is the reference magnetic field at magnetic axis.
The converged grid resolution for the global simulation at $n=7$ is $N_x \times N_y \times N_z \times N_{v_\|} \times N_{\mu } = 768 \times 1 \times 56 \times 48 \times 16$.
The local simulation uses $N_x = 64$ with the other dimensions unchanged.
% Here, $ a $ is the minor radius, $ \kappa _{A}=R/L_A $, and $ w_A $ is set to $ 0.15 $ by default.
% The converged grid resolutions for linear simulations are $ N_x \times N_y \times N_z \times N_{v_\|} \times N_{\mu } = 768 \times 1 \times 56 \times 48 \times 16 $ in the global simulation when $ n=7$.
% While local simulations require $ N_x = 64 $, and the others keep the same as the global resolutions.
% in local, $ N_x =64 $ while others keep the same as global are enough to converge the linear cases. % 

\begin{figure}[htbp]
  \begin{center}
    \begin{subfigure}{\linewidth}
      \centering
      \includegraphics[width=0.42\textwidth]{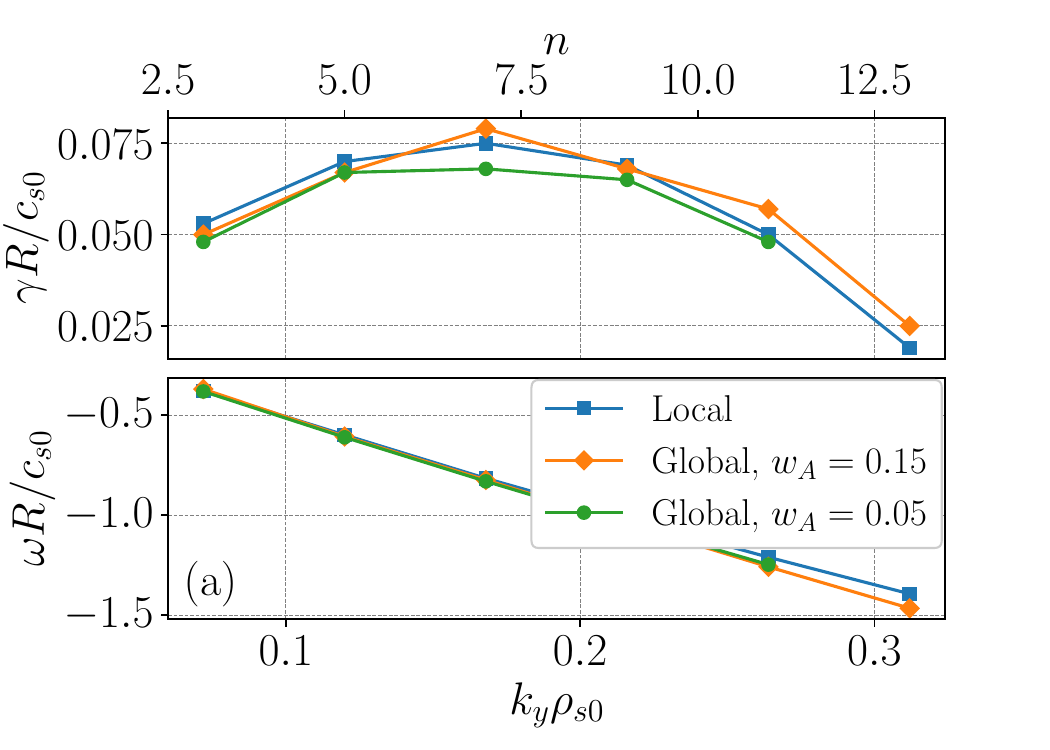}
      \includegraphics[width=0.40\textwidth]{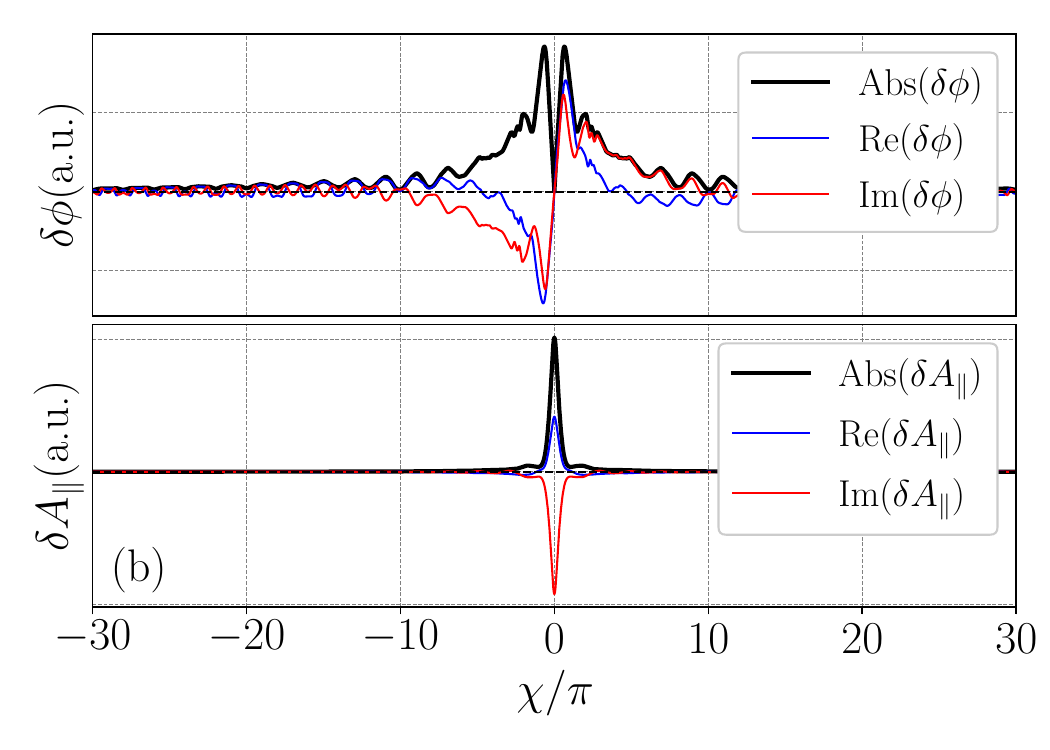}
    \end{subfigure}
    \begin{subfigure}{\linewidth}
      \centering
      \includegraphics[width=0.42\textwidth]{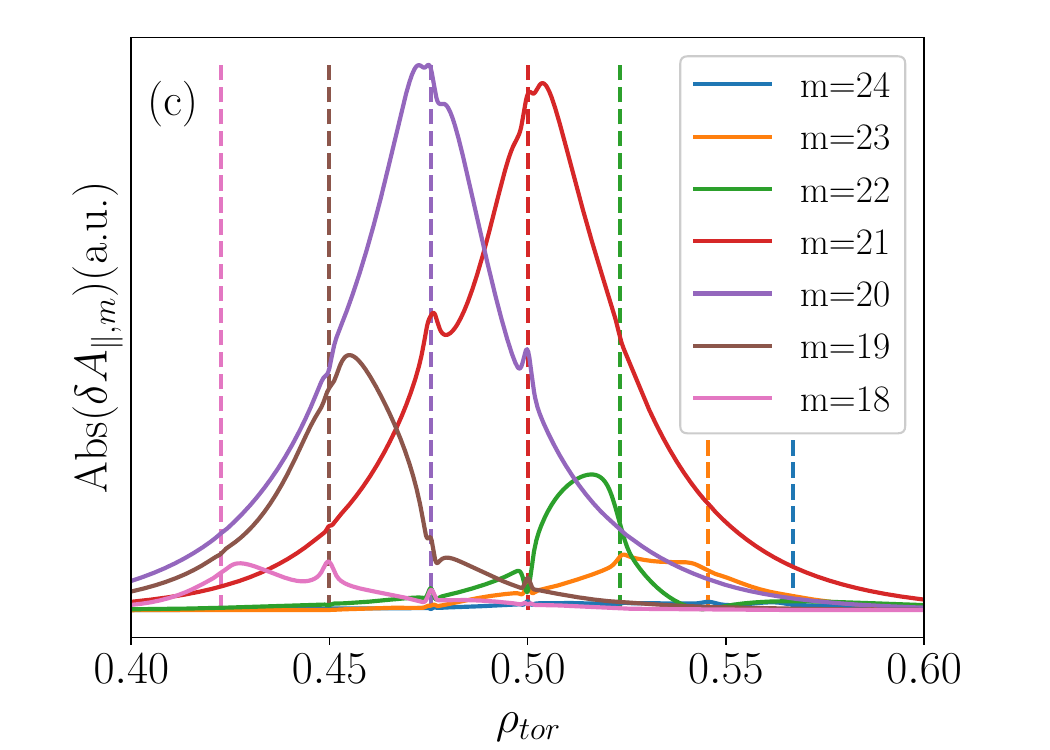}
      \includegraphics[width=0.42\textwidth]{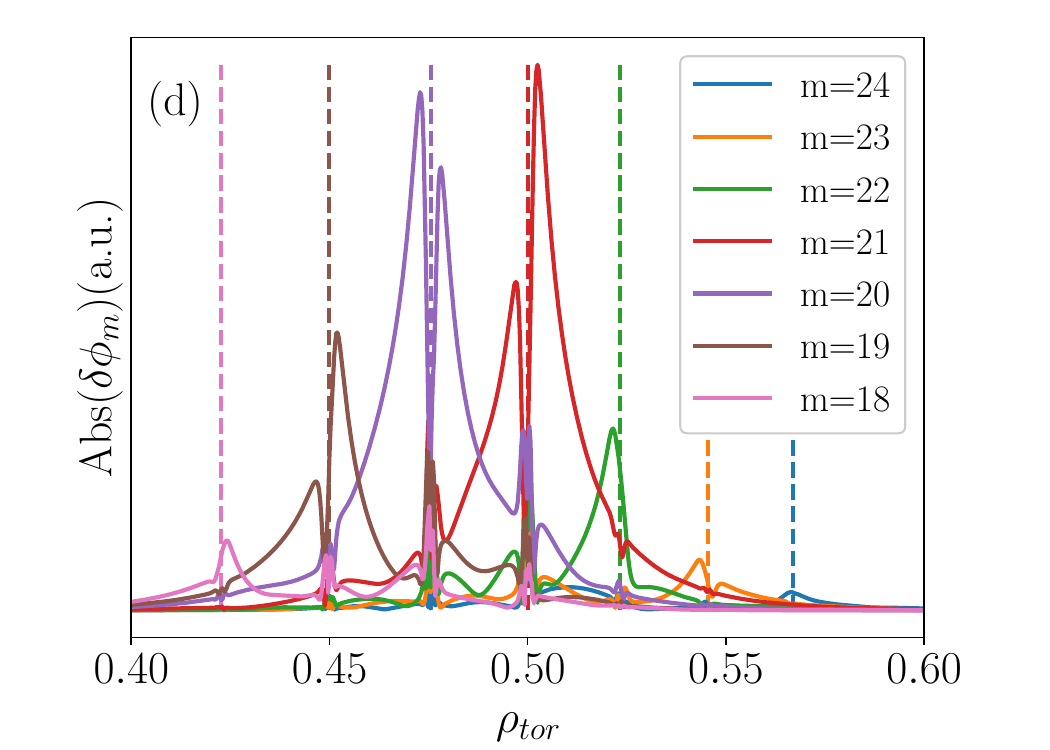}
    \end{subfigure}
    \begin{subfigure}{\linewidth}
      \centering
      \includegraphics[width=0.42\textwidth]{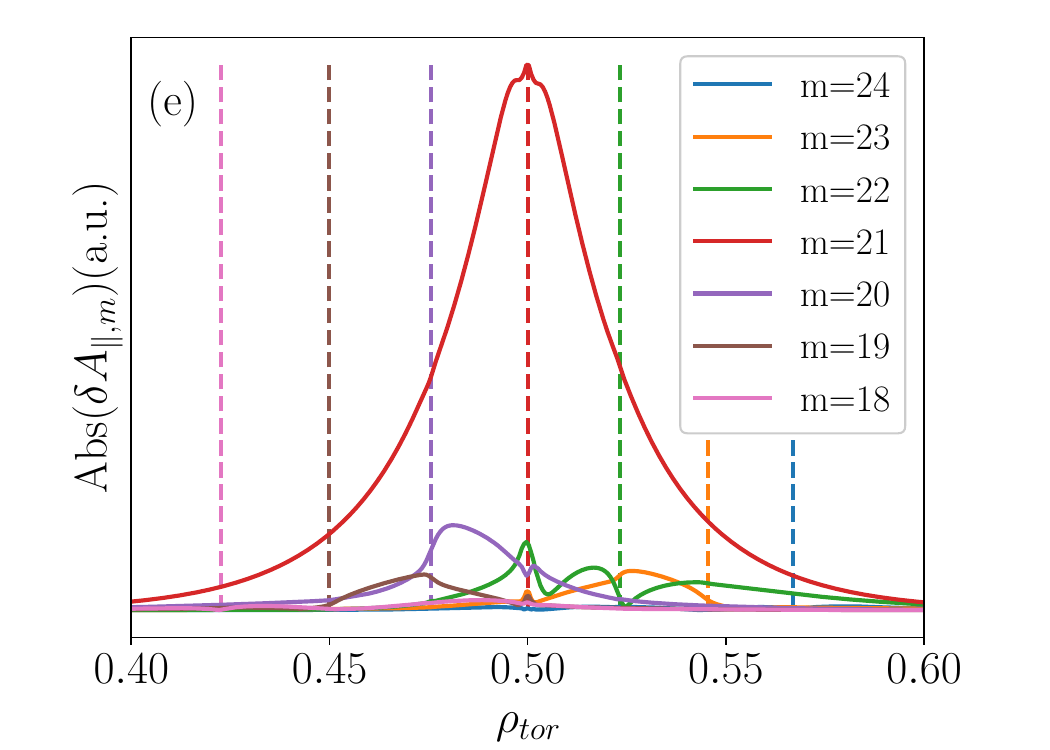}
      \includegraphics[width=0.42\textwidth]{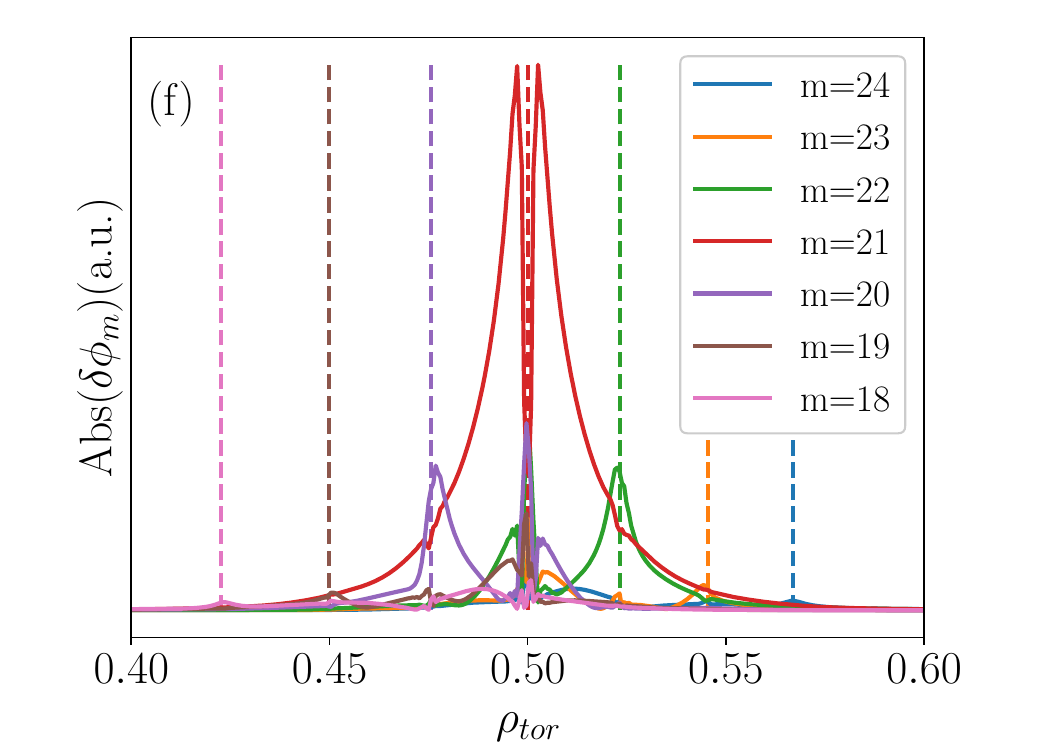}
    \end{subfigure}

  \end{center}
  \caption{(a) Comparisons of linear growth rates and frequencies of MTM versus $ k_y \rho_{s0} $ in local and global simulations. (b) Mode structures from local MTM simulation corresponding to $ n=7 $. (c)(d) Structures of harmonics of $ \delta A_\| $ and $ \delta \phi  $ in global MTM simulation corresponding to $ n=7 $. (e)(f) The same as (c) and (d) but $ w_A=0.05 $.}
  \label{fig:middle_gradient}
\end{figure}

Linear simulation results are shown in Fig. \ref{fig:middle_gradient}.
The growth rates and frequencies of local and global simulations are closely matched in Fig. \ref{fig:middle_gradient}(a).
The mode structures of local and global MTM simulations shown in Fig. \ref{fig:middle_gradient}(b-f) exhibit a standard tearing parity.
Figs. \ref{fig:middle_gradient}(c) and (d) present the global MTM structures of $w_A=0.15$ case with two dominant harmonics.
However, with a further reduction to $w_A=0.05$, only the harmonic corresponding to $m=nq_0$ becomes dominant, as shown in Figs. \ref{fig:middle_gradient}(e) and (f).
This suppression of harmonics at adjacent rational surfaces is attributed to the reduction of the gradients $ R/L_{A}(x) $ at their positions.
However, both the frequency and growth rate of the $ w_A=0.05 $ case align closely with the other cases, as illustrated in Fig. \ref{fig:middle_gradient}(a), which suggests that the MTMs are in fact solely excited at their respective MRSs.
% Although these modes may overlap, the effect of toroidal mode coupling is weak, as evidenced by small spikes of the dominant modes' $\delta A_{\|,m}$ at adjacent rational surfaces in Fig. \ref{fig:middle_gradient}(c).
Therefore, we refer to the MTMs in this regime as slab-like\cite{Chen2018}.

\begin{figure}[htbp]
  \begin{center}
    \begin{subfigure}{\linewidth}
      \centering
      \includegraphics[width=0.42\textwidth]{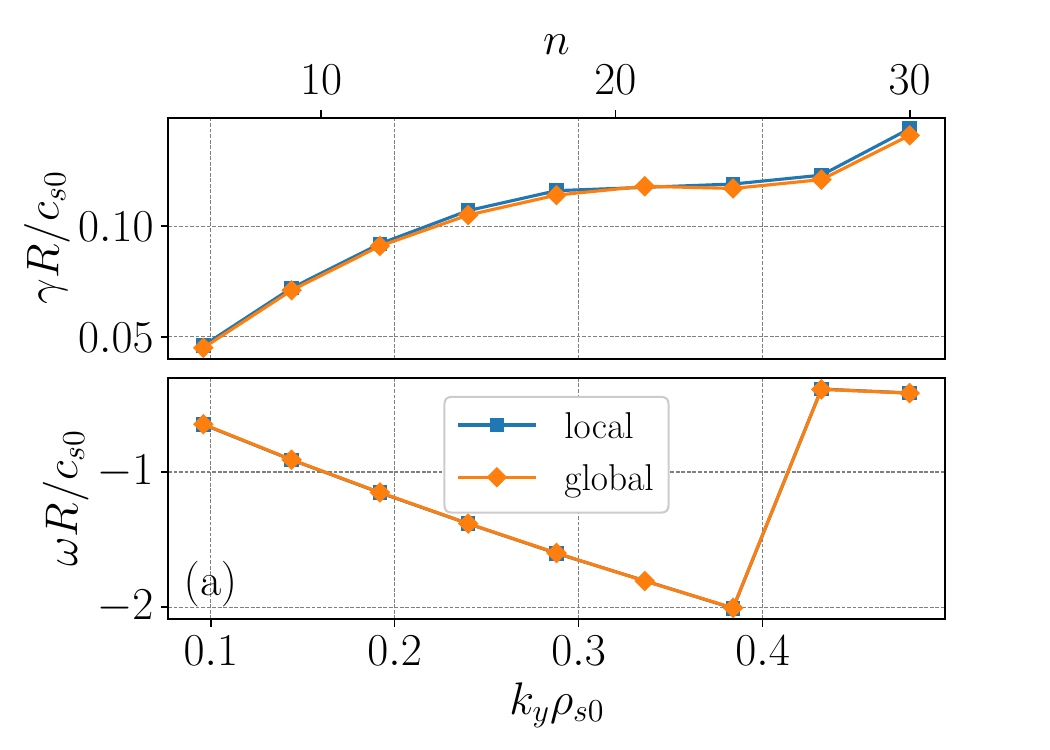}
      \includegraphics[width=0.38\textwidth]{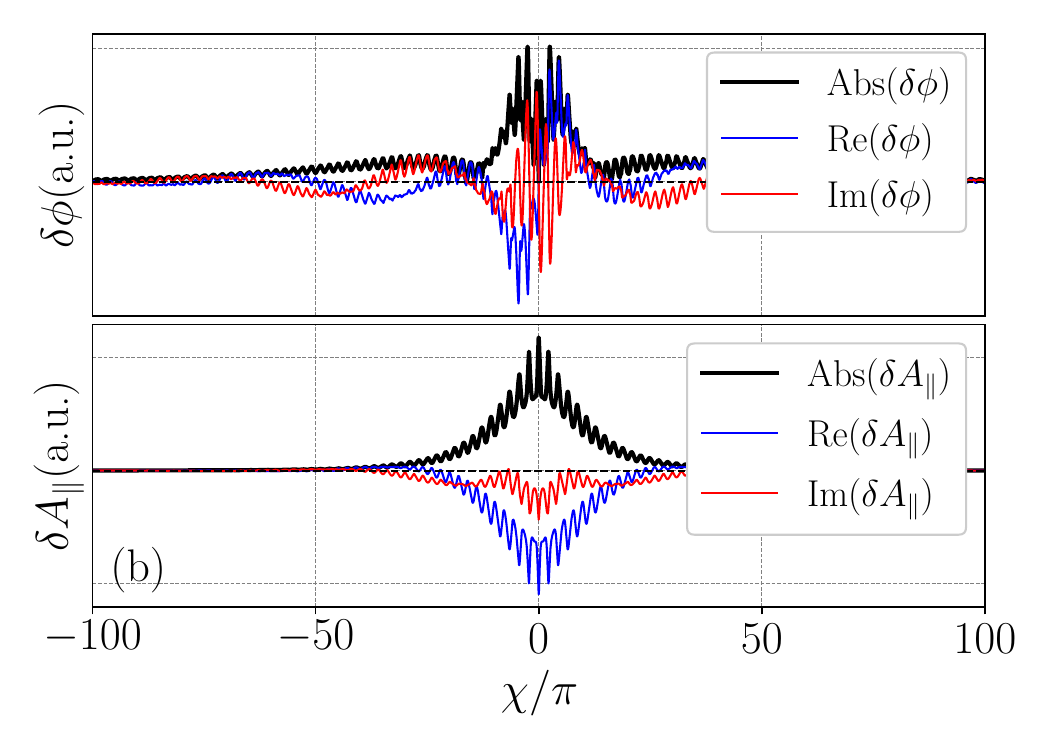}
    \end{subfigure}
    \begin{subfigure}{\linewidth}
      \centering
      \includegraphics[width=0.42\textwidth]{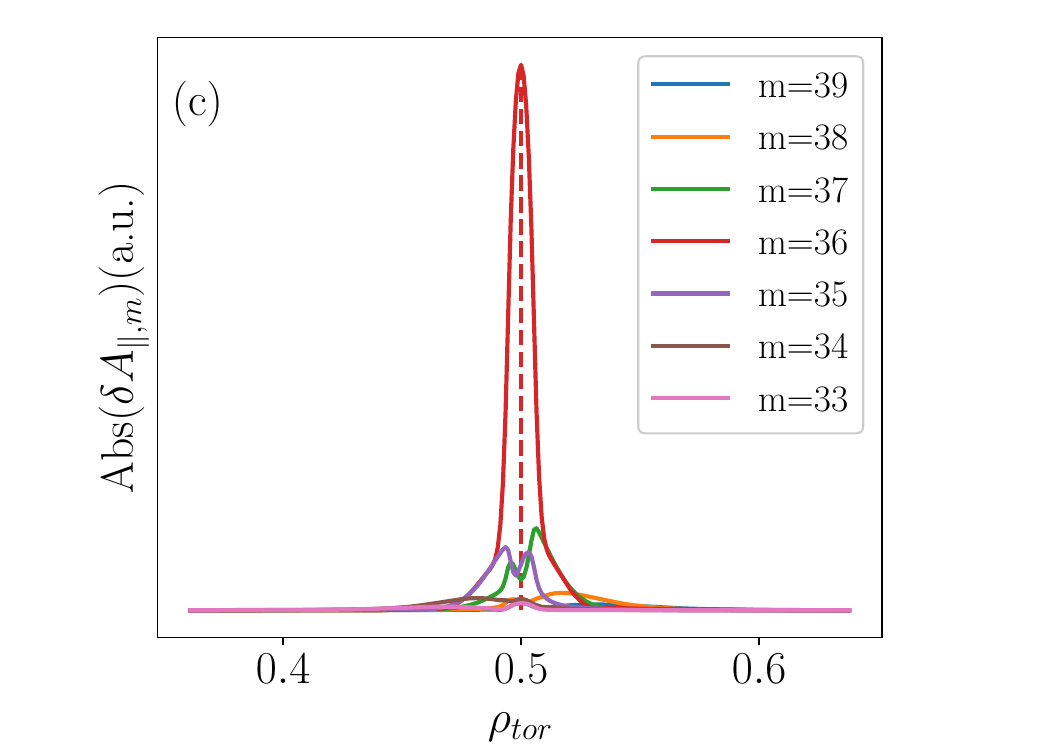}
      \includegraphics[width=0.42\textwidth]{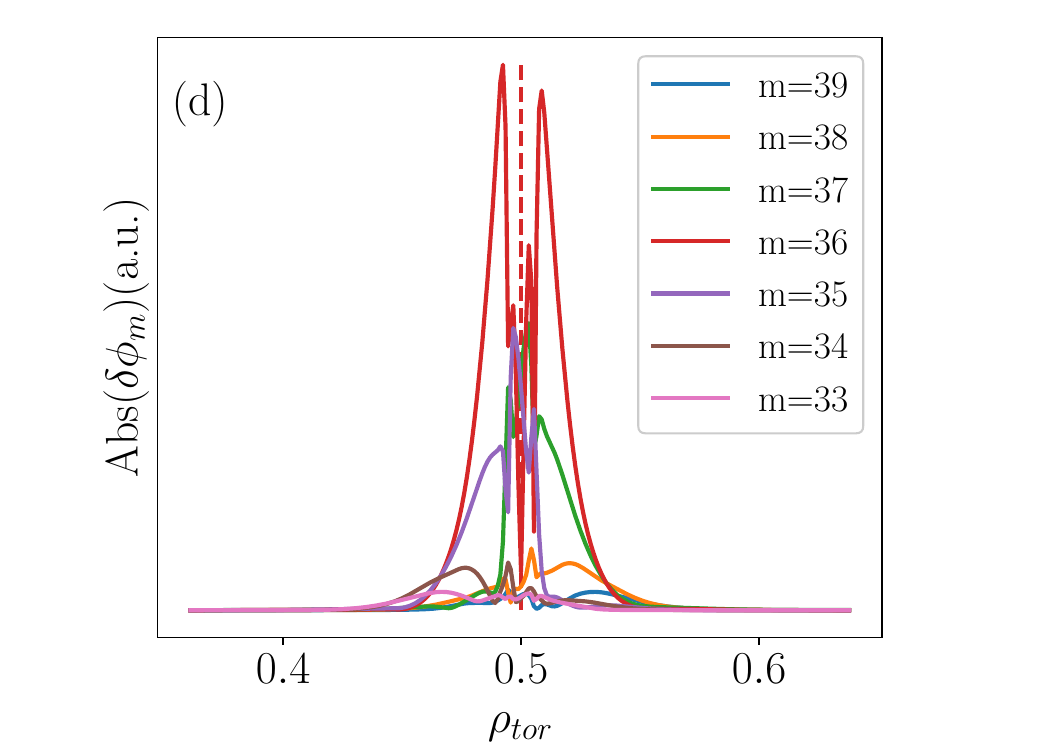}
    \end{subfigure}
  \end{center}
  \caption{(a) Comparisons of linear growth rates and frequencies of MTM versus $ k_y\rho_{s0} $ in local and global simulations in weak shear plasma. (b) Mode structures in local MTM simulation corresponding to $ n=18 $. (c)(d) Structures of harmonics of $ \delta A_\| $ and $ \delta \phi $ in global MTM simulation corresponding to $ n=18 $.}
  \label{fig:weakfreq}
\end{figure}

In weak magnetic shear cases, we adapt $ q= 1.8 + 0.4 x$ to have a weak shear plasma with $ \hat{s}=0.1 $ at $ x_0=0.5a $.
The inverse aspect ratio is increased to $ \epsilon = 0.3 $ and $ \kappa _{Te}=5 $,
while the other parameters keep as same as those in normal shear cases.
% Convergence tests require that $ N_x = 384 $ in global and $ N_x=128 $ in local.
The local and global simulations yield quantitatively consistent growth rates and frequencies, as shown in Fig. \ref{fig:weakfreq}(a).
The mode structures also exhibit the slab-like characteristic, with only a dominant harmonic localized at the MRS, as illustrated in Figs. \ref{fig:weakfreq}(c) and (d).
This corresponds to the significant extension of local mode structures along extended poloidal angle in Fig. \ref{fig:weakfreq}(b).
Moreover, the localization of the global mode structures renders the mode insensitive to the specific B.C.s or the domain size, $L_x$.
% The mode structures are observed to decay to zero at a distance exceeding half the RMS width from the mode's center, a region well within the radial boundaries. Consequently, the mode is insensitive to the specific B.C.s or the domain size, $L_x$.
% Their mode structures naturally converge to zero at a distance exceeding half the RMS width away from it, well before the radial boundaries.
% Consequently, the mode is insensitive to the outer B.C. or the $L_x$.
% , as confirmed by other scans where variations in these parameters have no essential impact on the results.
In fact, the local simulation here is equivalent to a global simulation constrained by $L_x=\Delta_m$ and periodic B.C.,
thus the observed consistency in frequencies and growth rates between local and global simulations is a direct consequence of the slab-like characteristic.

% It is widely recognized that in regions of weak $ \hat{s} $, toroidal ballooning parity modes such as the ITG mode are not appropriately described by local methods, e.g., the ballooning mode representation or flux-tube simulations \cite{Connor2004,Chen2018}.
% Therefore, under such conditions, global simulations serve as a proper research approach.
% Existing local simulations \cite{Jian2019,Jian2021} have predicted that MTM may exist in core regions of weak $ \hat{s} $, wherein a comparative analysis of the two simulation methods is necessary.

% It is widely recognized that in regions of weak $ \hat{s} $, toroidal ballooning parity modes such as the ITG mode are not appropriately described by local methods, e.g., the ballooning mode representation or flux-tube simulations \cite{Connor2004,Chen2018}.
% Besides, the profile variation of density and temperature gradients, as exemplified by the $ w_A $ scan in \cite{Villard2010}.

% This sensitivity arises because these modes extends radially across several MRSs through poloidal mode coupling.
% In contrast, the MTM is localized to the vicinity of its MRS and thus exhibits a strong resilience to profile variation.
The local approach, such as the ballooning mode representation \cite{Connor1979} and the flux-tube approach \cite{Beer1995}, is designed to simulate toroidal modes.
It requires a finite magnetic shear $\hat{s}$ and high toroidal mode number $n$ to ensure that the microscopic scale, $ \Delta_m $, is significantly narrower than the meso-scale \cite{Chen2018}.
This allows the formation of radially extended structures via toroidal mode coupling, thereby satisfying the quasi-translational invariance.
% However, the quasi-translational invariance is not a necessary condition for the applicability of local approaches.
However, for the slab-like MTMs studied in this work, the mode structures are solely excited and localized at the MRSs.
The local and global approaches become equivalent in these simulations, and the local simulations can capture the essential physics of MTM excitation. 
% For the slab-like MTMs studied in this work, which are localized at MRSs and with weak poloidal mode coupling, results of local and global simulations are consistent.
% However, slab-like MTMs occupy a distinct regime within the applicability of local methods.
% Specifically, these modes are localized near rational surfaces with weak poloidal mode coupling, exhibiting characteristics directly opposite to those of toroidal modes.
As a result, the slab-like characteristic manifests as a strong insensitivity of the mode against the variation of profiles.

\begin{figure}[htbp]
  \begin{center}
    \includegraphics[width=0.32\textwidth]{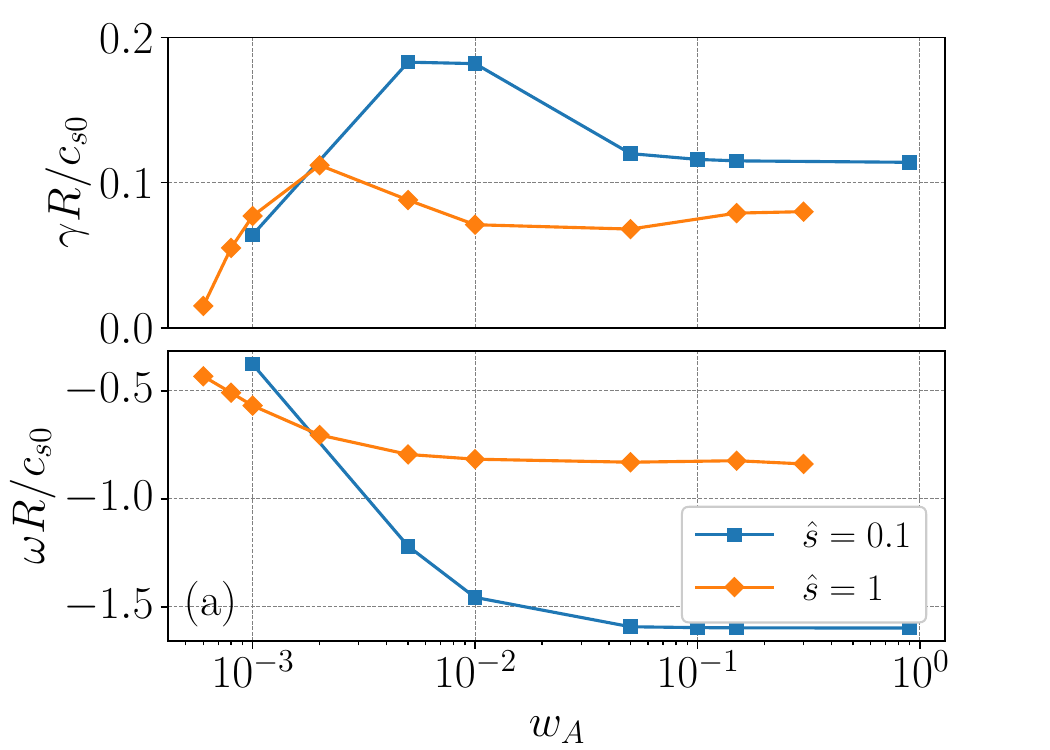}
    \includegraphics[width=0.32\textwidth]{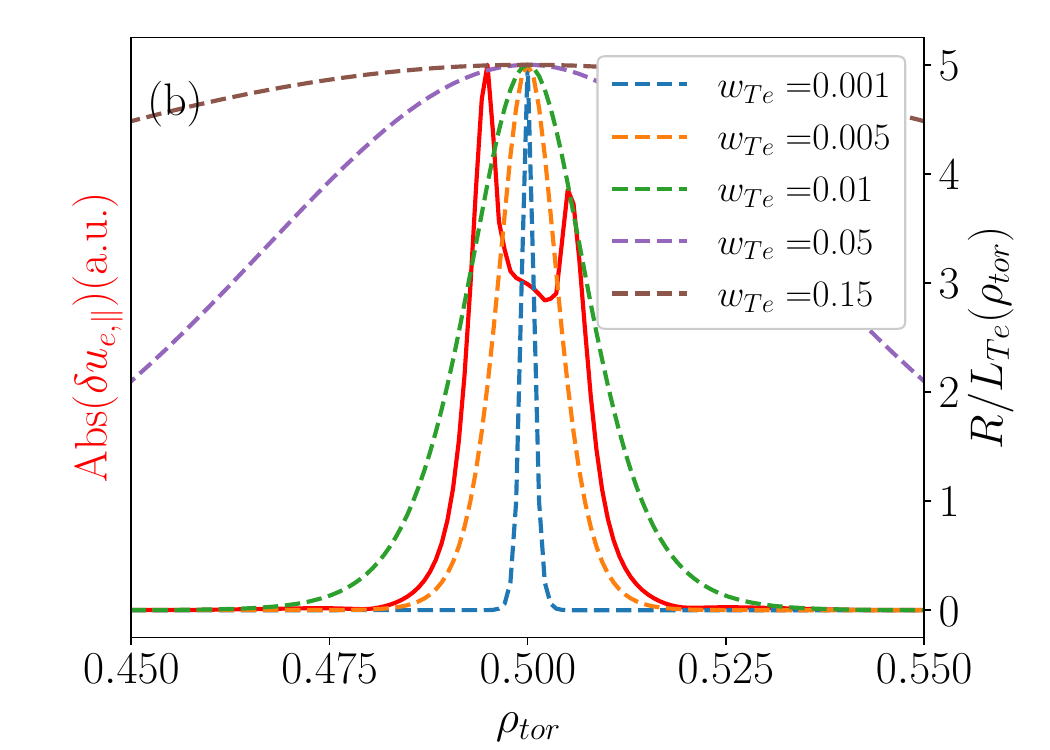}
    \includegraphics[width=0.32\textwidth]{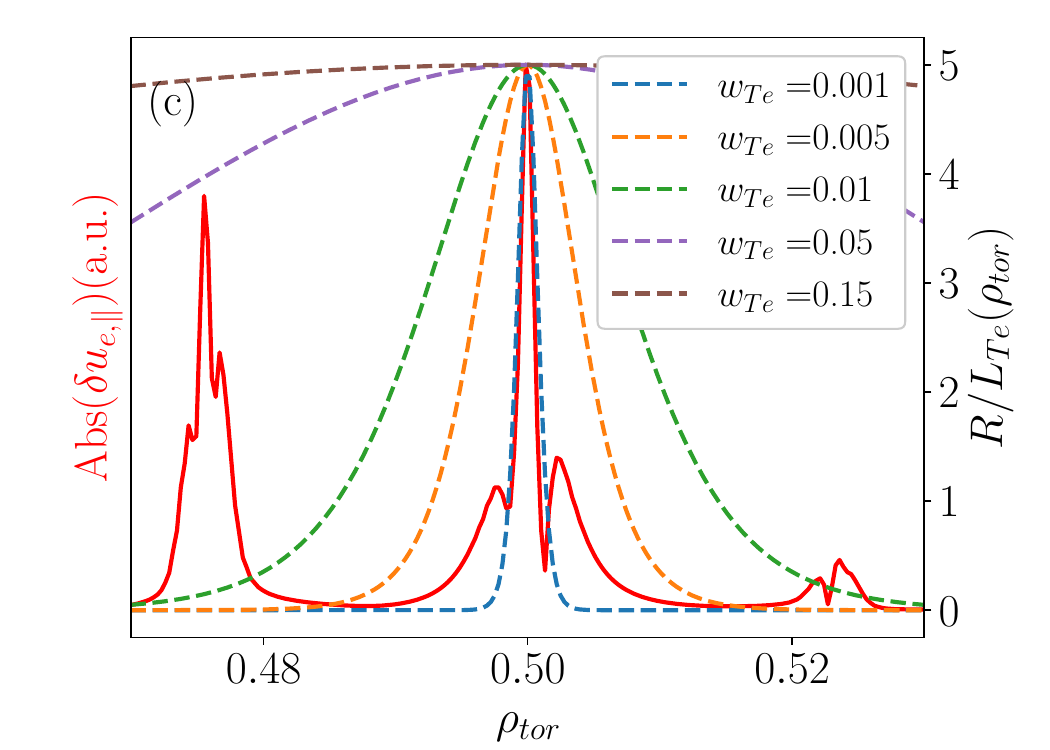}
  \end{center}
  \caption{(a) The growth rates and frequencies dependence on $ w_A $ in normal and weak magnetic shear cases.
    (b) Structure of the perturbed parallel electron velocity, $\delta u_{e,\|}$ (red solid line), obtained from a global simulation with $\hat{s}=0.1$ at $ x_0 $ and $w_A=0.15$, plotted alongside the $R/L_{Te}$ distribution (dashed lines) for various values of $w_{Te}$. Given that $|\delta u_{e,\|}| \gg |\delta u_{i,\|}|$, the mode structure of $\delta u_{e,\|}$ effectively represents the current layer structure.
  (c) The same as (b) but $ \hat{s}=1.0 $.}
  \label{fig:wascan}
\end{figure}

Moreover, the slab-like MTM is found to be influenced only by profile variations within a narrow scale of $\Delta_c$.
This conclusion is drawn from global simulations of $w_A$ dependence in Fig. \ref{fig:wascan}(a) and the corresponding comparison between current layer structures and $R/L_{Te}$ profiles in Figs. \ref{fig:wascan}(b) and (c). 
Consequently, for the slowly varying profiles typical of core plasmas, the discrepancy between local and global simulations of the slab-like MTM is expected to be insignificant.

\begin{figure}[htbp]
  \begin{center}
    \begin{subfigure}{\linewidth}
      \centering
      \includegraphics[width=0.42\textwidth]{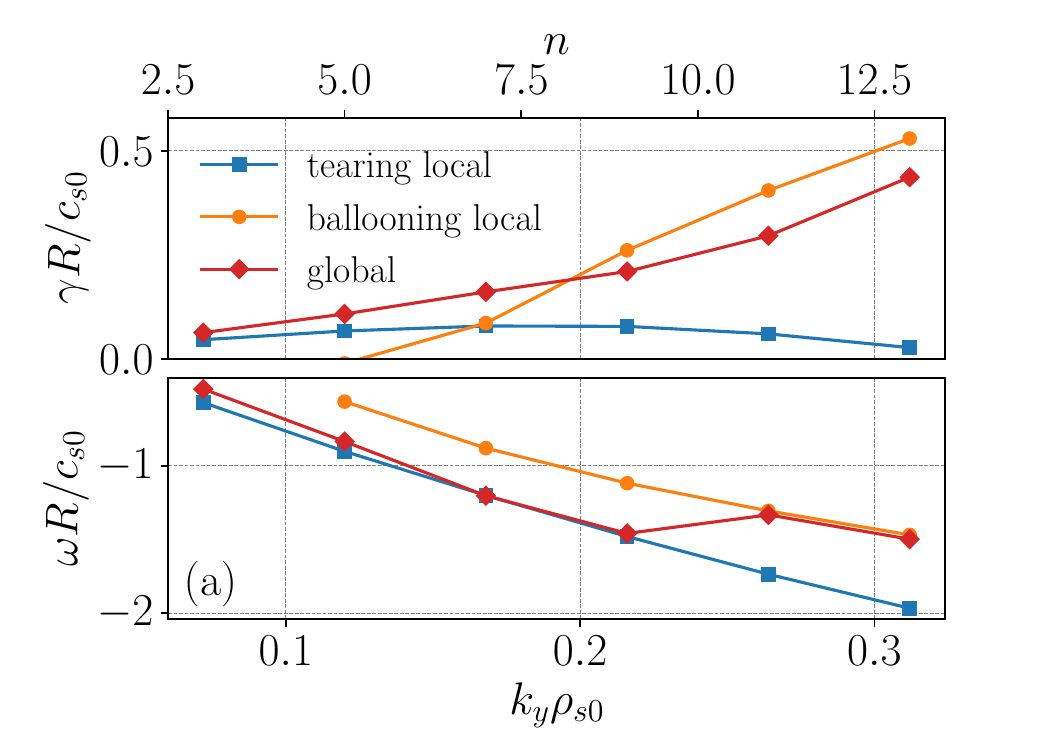}
      \includegraphics[width=0.42\textwidth]{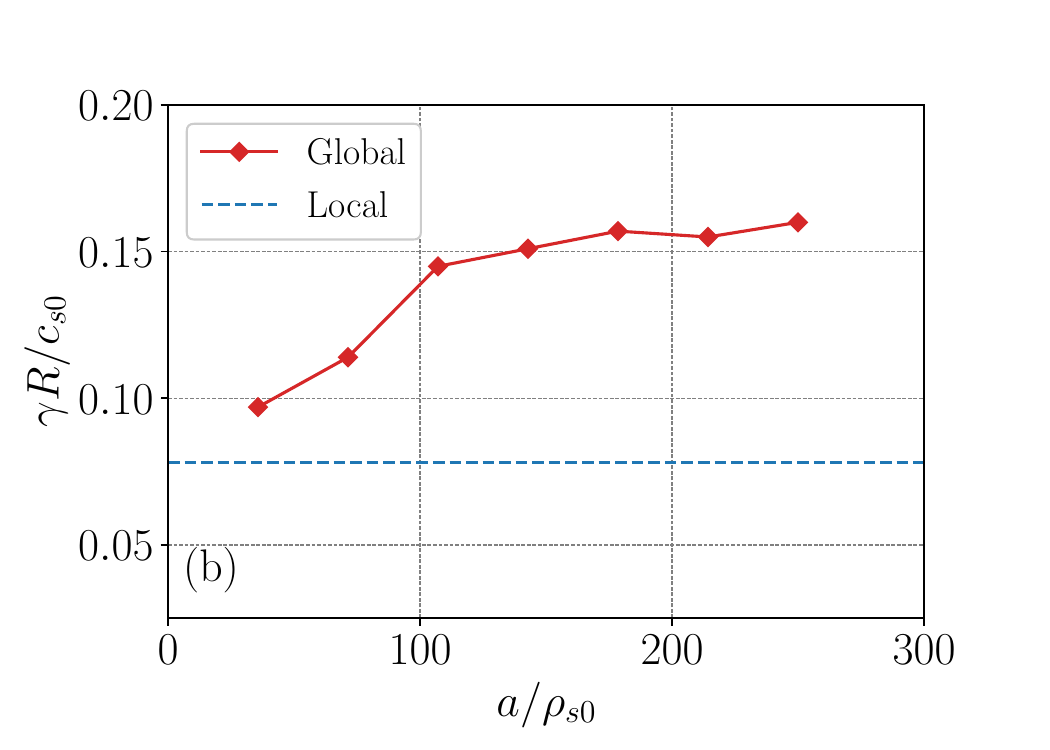}
    \end{subfigure}
    \begin{subfigure}{\linewidth}
      \centering
      \includegraphics[width=0.42\textwidth]{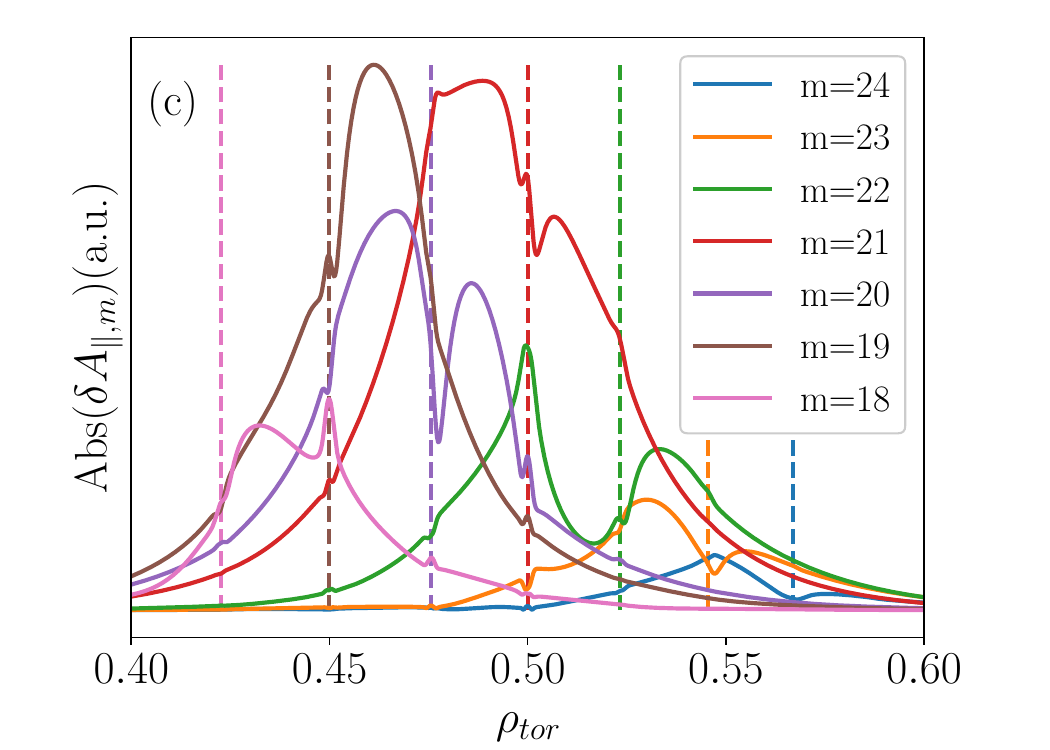}
      \includegraphics[width=0.42\textwidth]{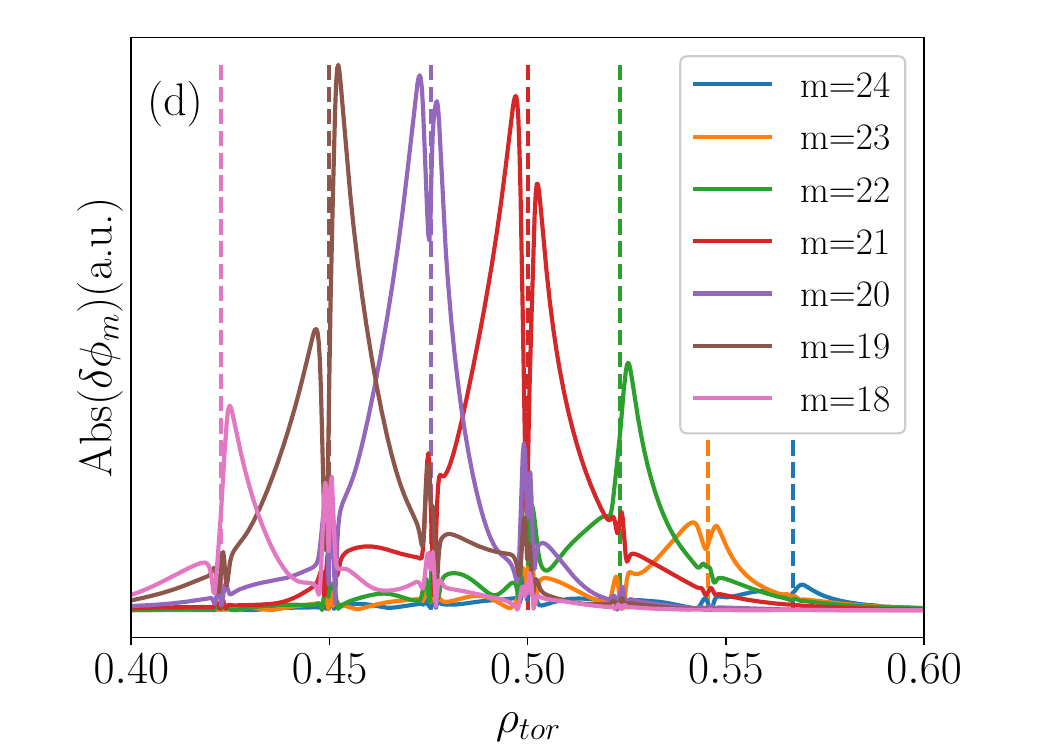}
    \end{subfigure}
  \end{center}
  \caption{(a) Comparisons of linear growth rates and frequencies of MTM versus $ k_y \rho _{s0} $ in local and global simulations, where $ R/L_{Te}=9 $. (b) Growth rates of global MTM simulation versus $ a/\rho_{s0} $ with fixed $ k_y\rho_{s0}=0.168 $. The dash line corresponds to the local results. (c)(d) Structures of harmonics of $ \delta A_\| $ and $ \delta \phi  $ in global MTM simulation corresponding to $ n=7 $.}
  \label{fig:stronger_gradient}
\end{figure}

However, MTMs are not always slab-like.
We find a type of MTM that has systematically higher growth rates in the global simulation compared to the local simulation, as shown in Fig. \ref{fig:stronger_gradient}(a) and (b), where $ R/L_{Te} $ is increased to $ 9 $ in the normal shear case.
The mode frequencies are nearly indistinguishable between the local MTM branch and the global one when $ n \le 9 $, which identifies the global modes with $ n \le 9 $ as MTMs.
As $ n $ increases, the dominant instability transitions into the trapped-electron mode (TEM).
These two branches in local simulations are isolated by introducing a module in GENE that filters the modes according to their different parities (see Appendix \ref{app:parity_filter} for details). 
Figs. \ref{fig:stronger_gradient}(c) and (d) illustrate that the harmonic structures in the global simulations deviate from the standard tearing parity shown in Fig. \ref{fig:middle_gradient} for $R/L_{Te}=5$. Instead, these structures exhibit a superposition of odd and even parities, a feature termed `parity mixing' in this study.

\begin{figure}[htbp]
  \begin{center}
    \begin{subfigure}{\linewidth}
      \centering
      \includegraphics[width=0.42\textwidth]{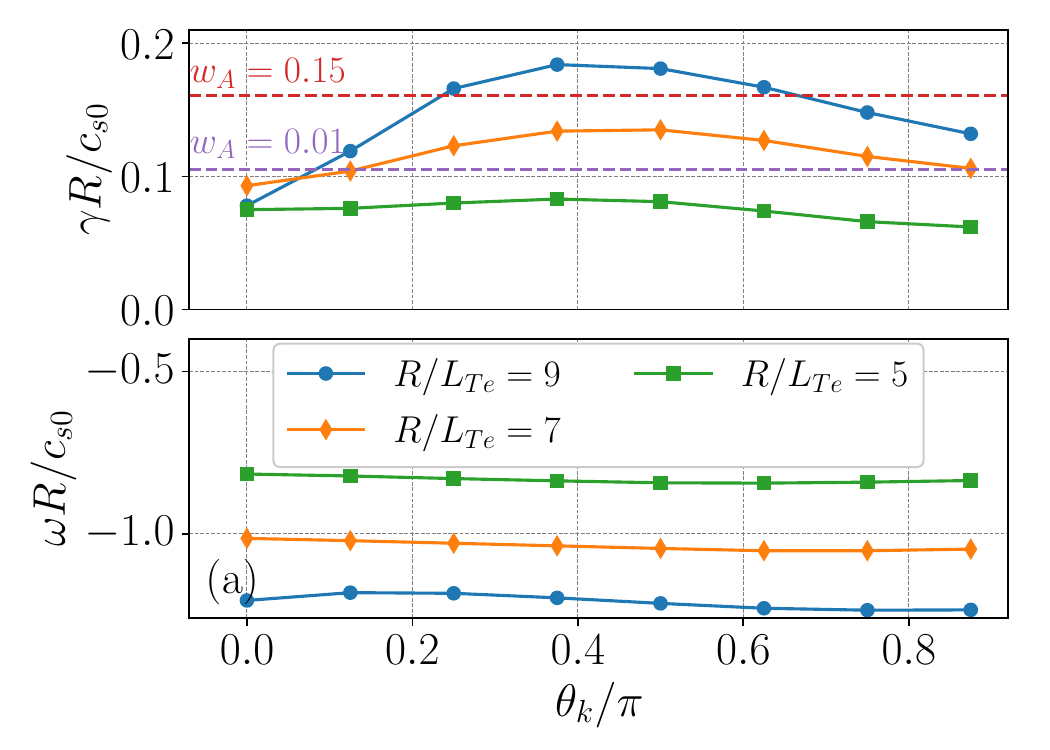}
      \includegraphics[width=0.42\textwidth]{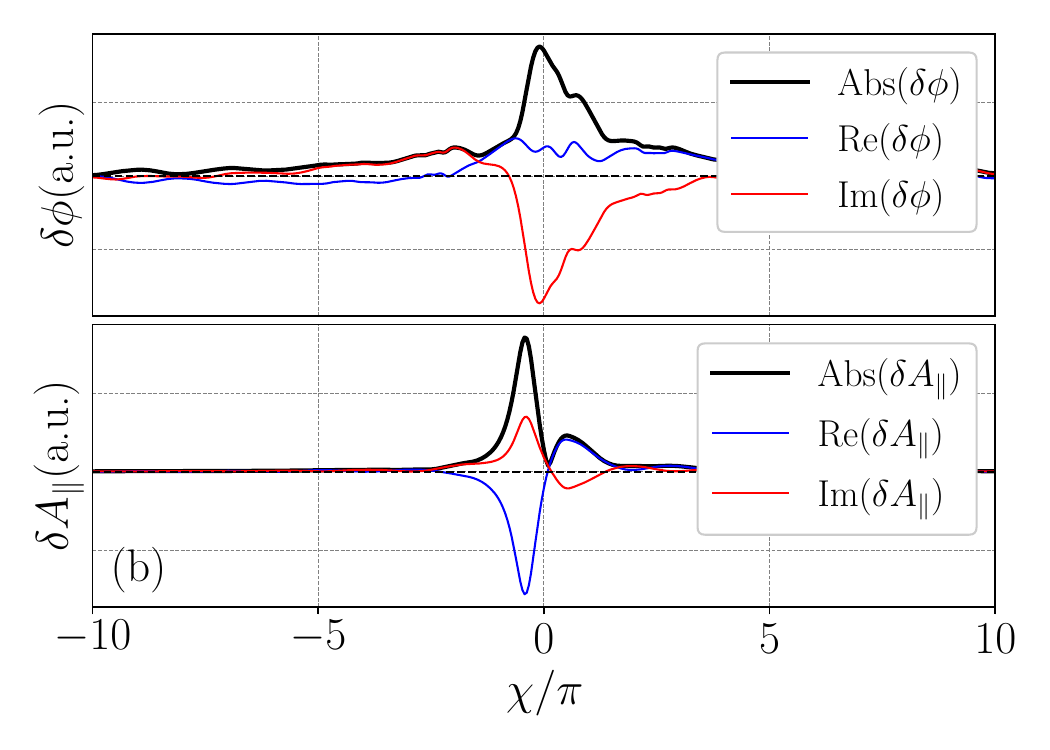}
      \includegraphics[width=0.42\textwidth]{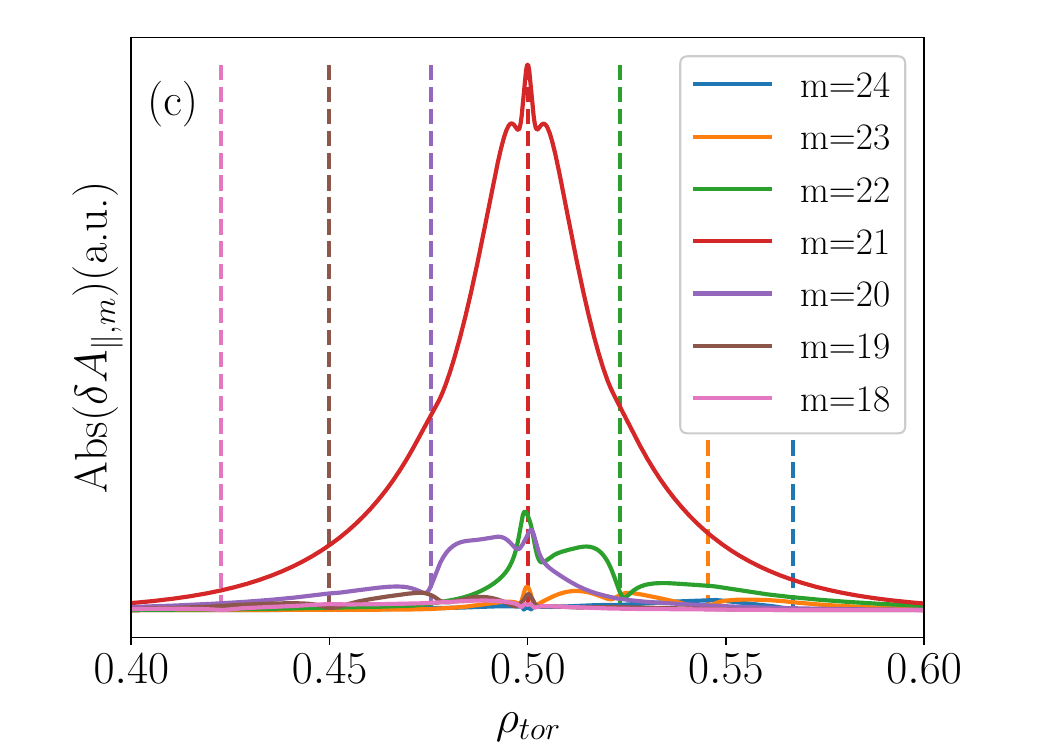}
      \includegraphics[width=0.42\textwidth]{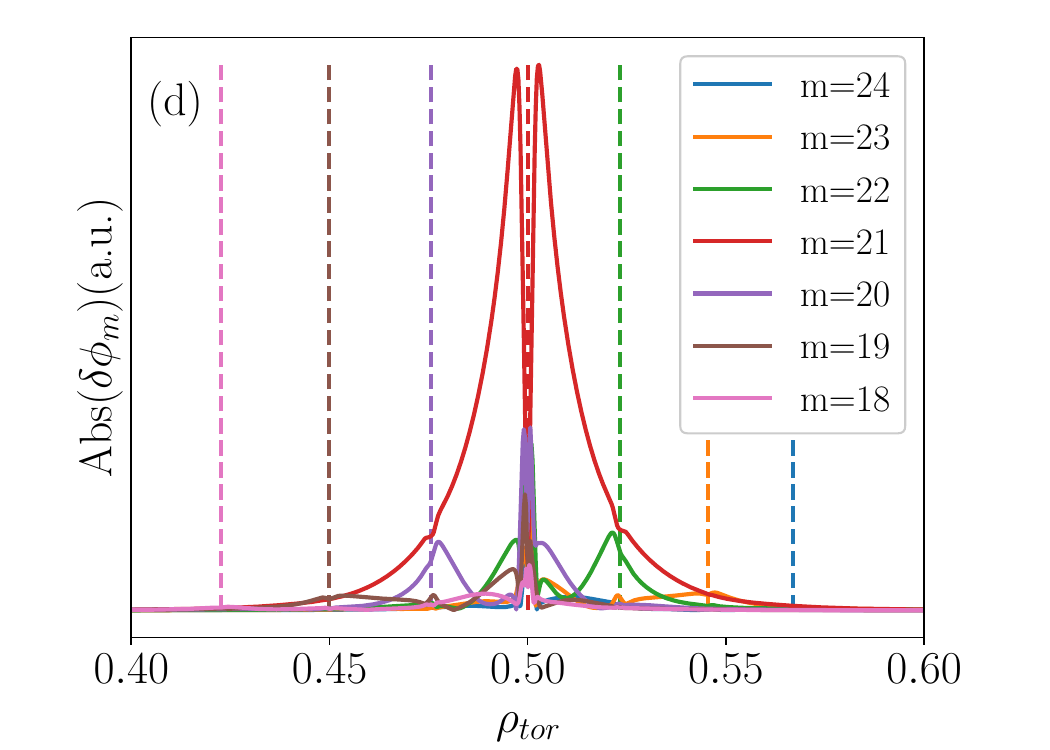}
    \end{subfigure}
  \end{center}
  \caption{(a) $ \theta _k $ scan of local simulations corresponding to $ n=7 $ with different gradient. The growth rates of global simulations with $ w_A=0.15 $ and $ w_A=0.01 $ are labeled with red and purple dashed lines, respectively. (b) Mode structures of MTM with $ R/L_{Te} = 9 $, $ \theta _k = 0.375 \pi $. (c)(d) Structures of harmonics of $ \delta A_\| $ and $ \delta \phi  $ in global MTM simulation corresponding to $ n=7 $ and $ w_A=0.01 $.}
  \label{fig:kxscan}
\end{figure}

Within the framework of the ballooning mode representation, the deviation of the global results from the local ones in Fig. \ref{fig:stronger_gradient}(a) stems from the finite $\theta_k$ effect \cite{Dewar1997},
which was omitted in the local simulations presented earlier.
Its impact is investigated in Fig. \ref{fig:kxscan}(a) through scans with different temperature gradients.
It is observed that for $R/L_{Te} = 9$, the growth rate at $\theta_k \simeq 0.4 \pi$ exceeds that of the mode at $\theta_k = 0$ significantly. % and is marginally higher than the global simulation result.
%TODO: a space after figure and cite
For finite $\theta_k$, the parity of the corresponding mode structure becomes indistinguishable, as shown in figure\ref{fig:kxscan}(b).
This is consistent with the parity mixing in Figs. \ref{fig:stronger_gradient}(c) and (d).
% The local structures in Fig. \ref{fig:kxscan}(b), which approach an even-parity-like mode, are consistent with the parity mixing phenomenon in figures \ref{fig:stronger_gradient}(c)(d).
Moreover, reducing $w_A$ suppresses the parity mixing in global simulations (see Figs. \ref{fig:kxscan}(c) and (d)), causing the growth rate to approach the zero-$\theta_k$ local result, as shown in Fig. \ref{fig:kxscan}(a).
This indicates that the toroidal coupling between harmonics is important in destabilizing this kind of MTM.
Thus we identify the MTM as a toroidal mode. 
In addition, the independence of the growth rate on $\theta_k$ at $R/L_{Te}=5$ in Fig. \ref{fig:kxscan}(a) suggests that the slab-like MTM is unaffected by radial envelope modulation, further confirming the absence of toroidal coupling between its harmonics.

Further investigation reveals that the MTM destabilized by finite $\theta_k$ originates from the trapped electron dynamics.
Since the bounce frequency of trapped electrons satisfies $ \omega _b \gg \omega $, the bounce-averaged electromagnetic drift-kinetic equation \cite{Chen1991,Chen2018en,Yao2024,Chen2025} in ballooning mode representation is:
\begin{align}
	\delta K_{te,j}^{(0)}\left( \kappa ^2,v^2\right) = - \frac{\omega -\omega _{*e}^{T}}{\omega -\left\langle \omega _{de } \right\rangle_T}\frac{eF_{e0}}{T_e}\left\langle \delta S_1 \right\rangle_{j,T},  j \in \mathbb{Z},
	\label{eq:bak}
\end{align}
here,
$ \delta K_{te,j}^{(0)}$ is the zero order bounce-averaged ($ \mathcal{O}(\omega /\omega _b) $) nonadiabatic distribution of trapped electrons at the $ j $th period of extended poloidal angle,
$ \delta K_{te,j}^{(0)}$ is constant in $ 2 j \pi - \eta _0<\eta <2 j\pi +\eta _0 $,
$ \delta S_1=\delta \phi -(1- \omega _{de}  /\omega)\delta \psi $,
$ \delta \psi = \omega \delta A_\|/c k_\| $,
$ \left\langle \omega _{de} \right\rangle_T $ is the precession frequency of trapped electrons.
% $ \delta B_\| $ and the finite Larmor radius effect of electrons are neglected for MTM research.
The bounce average operator is defined as
\begin{equation}
	\left\langle A \right\rangle_{j,T}= \frac{1}{4 K\left( \kappa  \right)}\int _{2j \pi -\eta _0\left( \kappa  \right)}^{2j \pi + \eta _0\left( \kappa  \right)} \frac{A\left( \alpha  \right)}{\sqrt{\kappa ^2-\sin ^2 \frac{\alpha }{2}} }\mathrm{d}\alpha,
	\label{eq:bcav}
\end{equation}
where $ \kappa ^2 $ is the pitch angle variable,
$ \eta _0= 2 \arcsin |\kappa | $.

\begin{figure}[htbp]
  \begin{center}
    \includegraphics[width=0.42\textwidth]{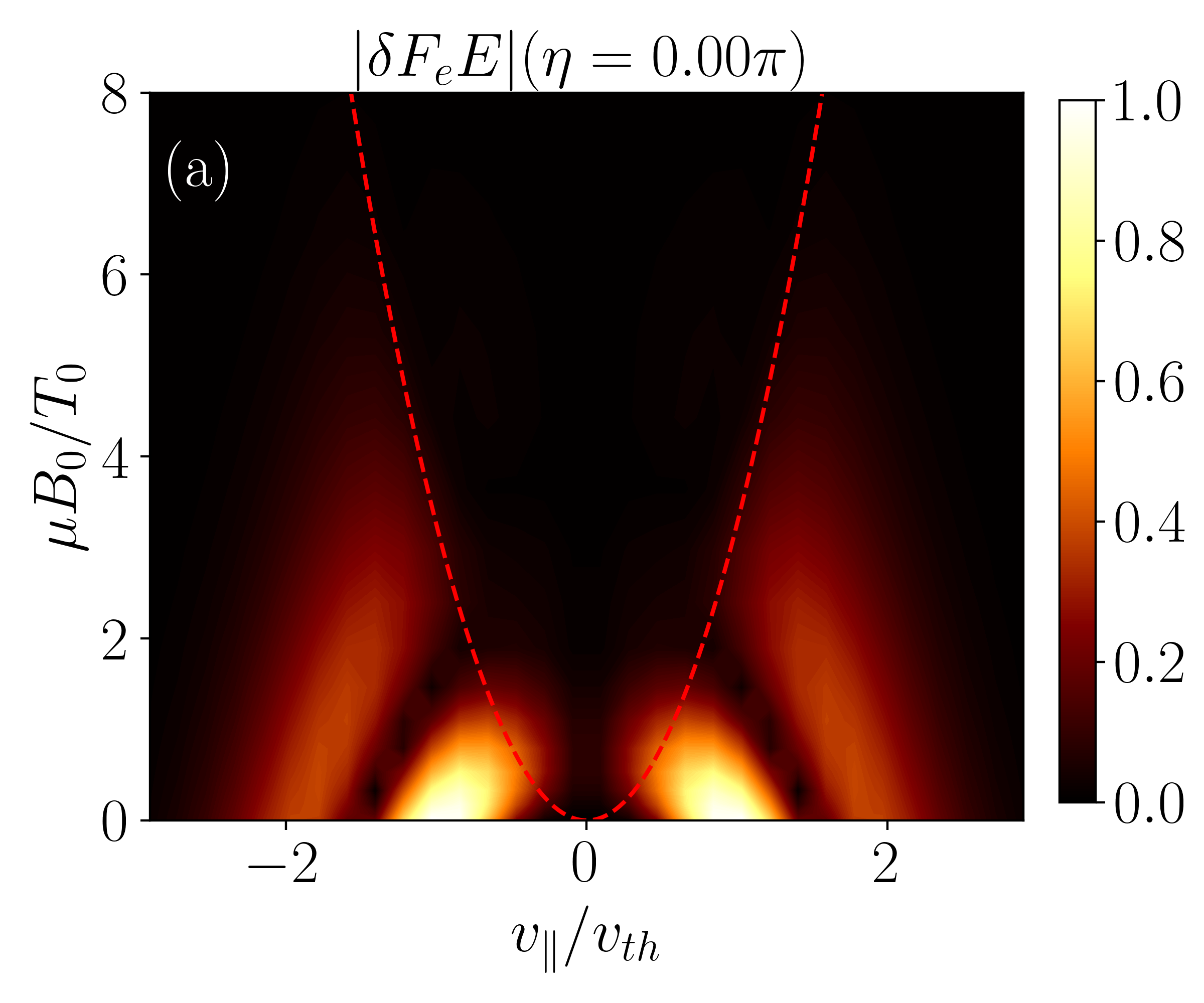}
    \includegraphics[width=0.42\textwidth]{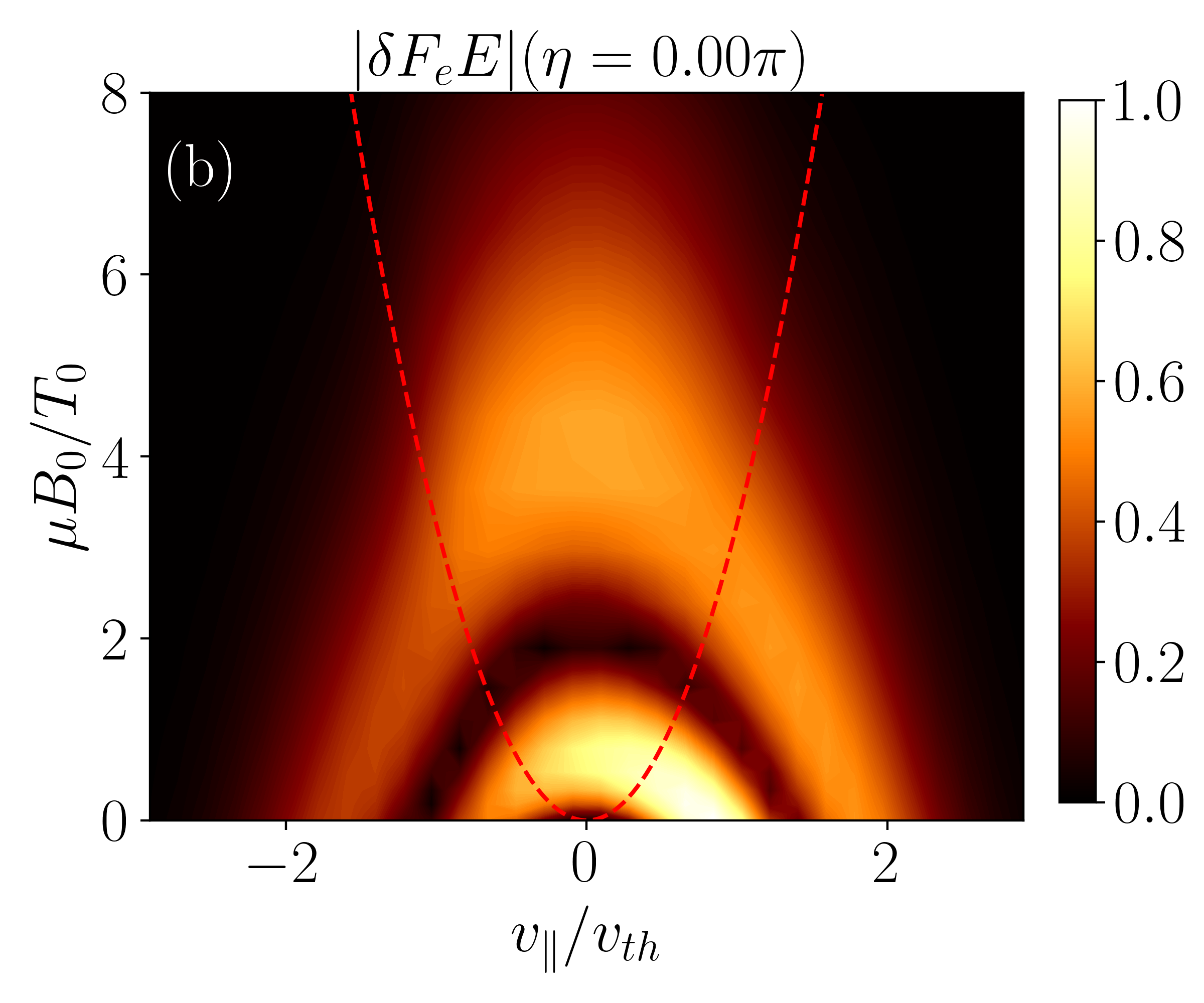}
  \end{center}
  \caption{The response of electrons to the MTM, characterized by the amplitude of energy exchange $ |\delta F_e E| $. The parameters correspond to the case of $ n=7 $, $ R/L_{Te}=9 $ in Fig. \ref{fig:kxscan}(a). The red dashed line indicates the trapped-passing boundary. The ballooning angles are (a) $ \theta _k=0 $ and (b) $ \theta _k=0.375\pi$.}
  \label{fig:vsp-g9n7th3}
\end{figure}

For the MTM with $ \theta _k=0 $, $ \delta S_1\left( \eta  \right) $ is an odd function. Thus $ \delta K_{te,0}^{(0)} \propto \left\langle \delta S_1 \right\rangle_{j,T}$ is exactly equal to $ 0 $, while terms for $ j\neq 0 $ remain finite.
Noting that $ \delta S_1 $ peaks at $ j=0 $ in general, thus the cancellation induced by bounce dynamics and the odd parity mode structure significantly diminishes the contribution of trapped electrons to the mode. 
In local simulations, Fig. \ref{fig:vsp-g9n7th3} (a) shows the weak response of trapped electrons in $ \left( v_\|,\mu \right) $ phase space at $ \eta = 0 $ with $ \theta _k =0 $.
However, the finite $ \theta _k $ breaks the odd parity as shown in Fig. \ref{fig:kxscan}(b), accompanied by a significant response of trapped electrons is observed in Fig. \ref{fig:vsp-g9n7th3} (b).
This is similar to the effects of $ \theta _k $ on TEMs \cite{Chen2018a}.
It is worth noting that MTMs are typically considered stable in the tokamak core\cite{Connor1990}. However, our results above demonstrate that trapped electron dynamics can introduce a significant destabilizing mechanism for MTM in core plasmas.  Via the parity mixing effect at a finite $\theta_{k}$, this mechanism renders their growth rates much higher than predicted by conventional theories.
% Therefore, trapped electron dynamics may provide an additional destabilizing mechanism for MTMs in core plasmas, as demonstrated in Figs. \ref{fig:stronger_gradient} (a) and \ref{fig:kxscan} (a), making them substantially more competitive than predicted by conventional MTM theory, in which MTMs are typically regarded as weak instabilities relative to reactive modes such as TEMs.

\section{MTM simulations in pedestal plasmas}
\label{sec:Strong Gradient}
\begin{figure}[htbp]
  \begin{center}
    \includegraphics[width=0.3\textwidth]{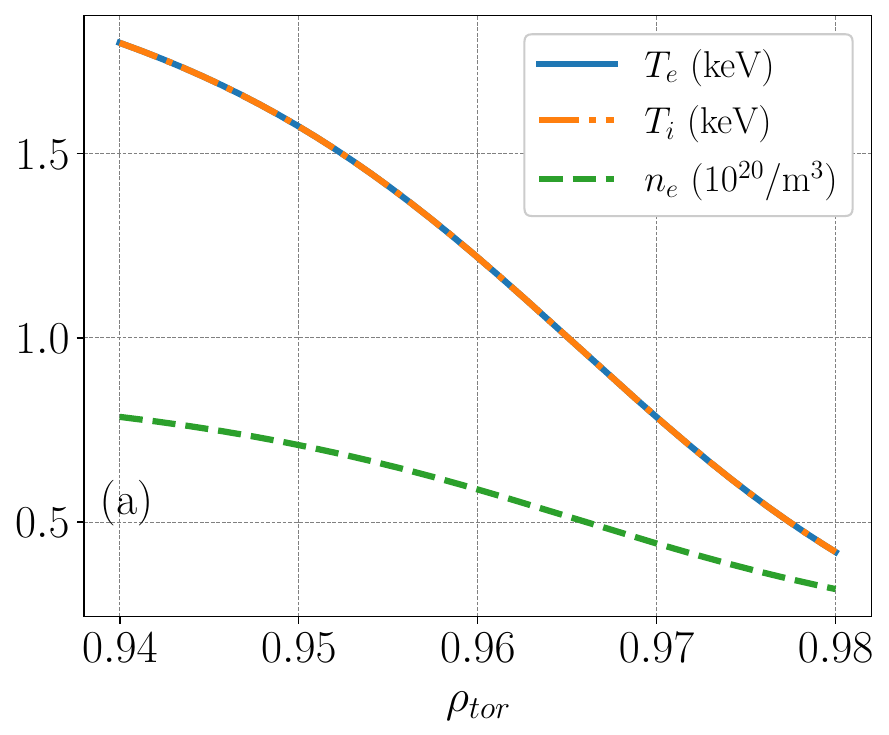}
    \includegraphics[width=0.3\textwidth]{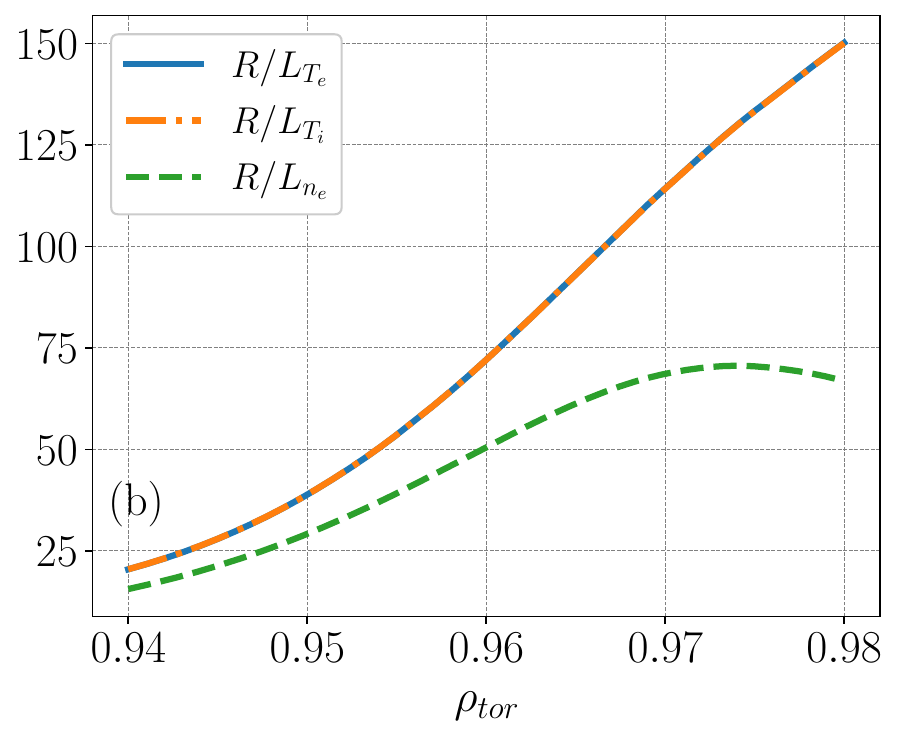}
    \includegraphics[width=0.33\textwidth]{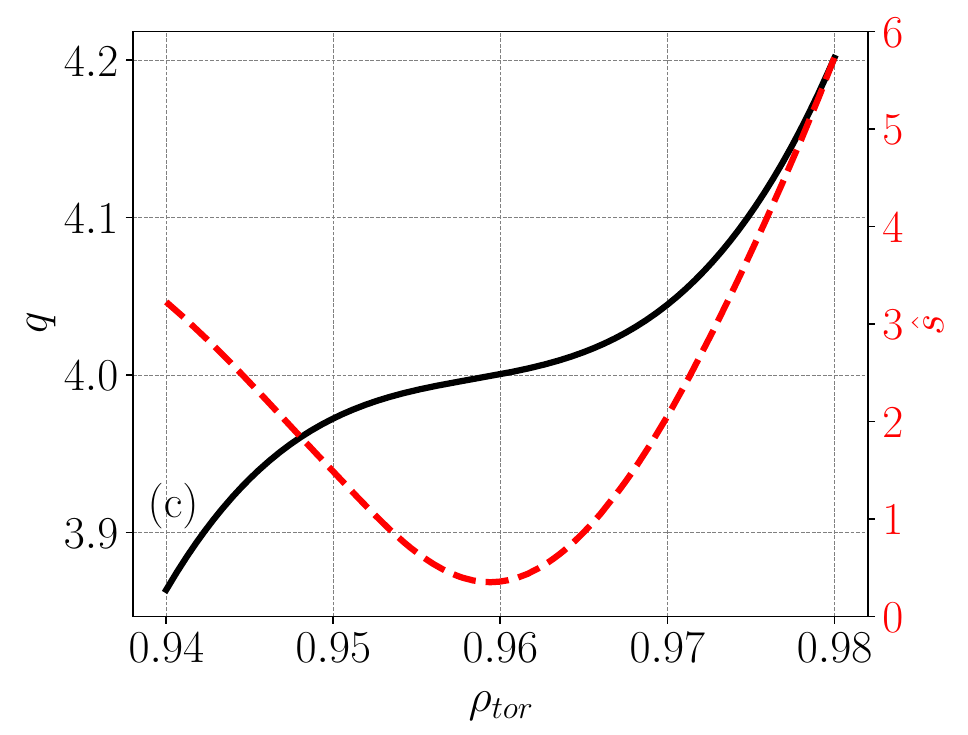}
  \end{center}
  \caption{Profiles of the pedestal simulation.}
  \label{fig:ped_profile}
\end{figure}

MTMs are observed in the pedestal region of tokamaks\cite{Hatch2021,Hassan2021,Halfmoon2022,Curie2022,Chen2023a,Jian2024}, an area characterized by the strong plasma non-uniformity.
Taking the advantage of GENE's MHD equilibrium interface,
we utilize an equilibrium generated from an integrated simulation using OMFIT \cite{Meneghini2015}, which provides the self-consistent profiles and geometry required for the global simulation of the pedestal region.
The profiles of the concerned region are shown in Fig. \ref{fig:ped_profile}.
The midpoint for global simulations and the reference point for local simulations are set at $ \rho _{tor}=0.96 $ with $ q\simeq 4 $.

\begin{figure}[htbp]
  \begin{center}
    \includegraphics[width=0.4\textwidth]{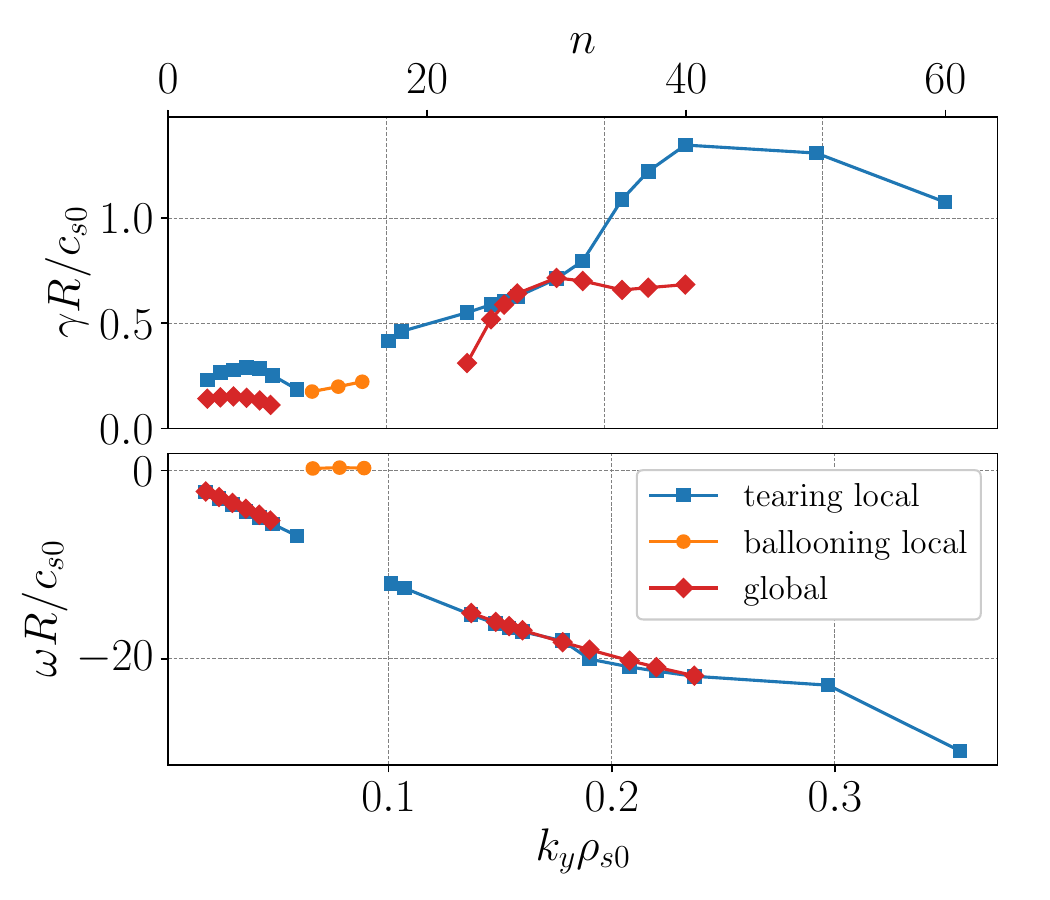}
  \end{center}
  \caption{Comparison of local and global growth rates and frequencies versus $ k_y\rho_{s0} $ in the pedestal region.}
  \label{fig:ped_freq}
\end{figure}

Fig. \ref{fig:ped_freq} presents the comparison between local and global simulation results,
where global simulations mostly exhibit lower growth rates than their local counterparts.
Nevertheless, both simulation approaches consistently identify two distinct MTM branches in the low-$ n $ ($ n\leq 10 $) and high-$ n $ ($ n\geq 20 $) regimes, separated by an intermediate branch dominated by ballooning parity modes.
Unlike the results in Section \ref{sec:Normal Gradient}, no finite $\theta_k$ destabilization is observed in the local simulations (data not shown here).

It is found that the current layer of the MTM in pedestal region is much wider than that in the core region.
This has substantial influences on both low-$ n $, and, especially, high-$ n $ modes, leading to discrepancies between local and global results.
$ \Delta _c $ can be estimated by
\begin{equation}
  \frac{\Delta _c}{\rho _{s0}}\simeq \frac{q}{\hat{s}}\frac{\omega /\left( c_{s0}/R \right)}{k_y \rho _{s0}}\sqrt{\frac{m_e}{m_H}}.
  \label{eq:cl}
\end{equation}
For MTMs, $ \omega \simeq \omega _{*p e}=k_y\rho_{s0} \left( c_{s0}/R \right)\left( R/L_{ne} + R/L_{Te}\right) $,
substituting this expression into equation \eqref{eq:cl} yields
\begin{equation}
  \frac{\Delta _c}{\rho _{s0}} \simeq \frac{q}{\hat{s}}\sqrt{\frac{m_e}{m_H}} \left( \frac{R}{L_{ne}}+\frac{R}{L_{Te}} \right).
  \label{eq:cl2}
\end{equation}

\begin{figure}[htbp]
  \begin{center}
    \includegraphics[width=0.32\textwidth]{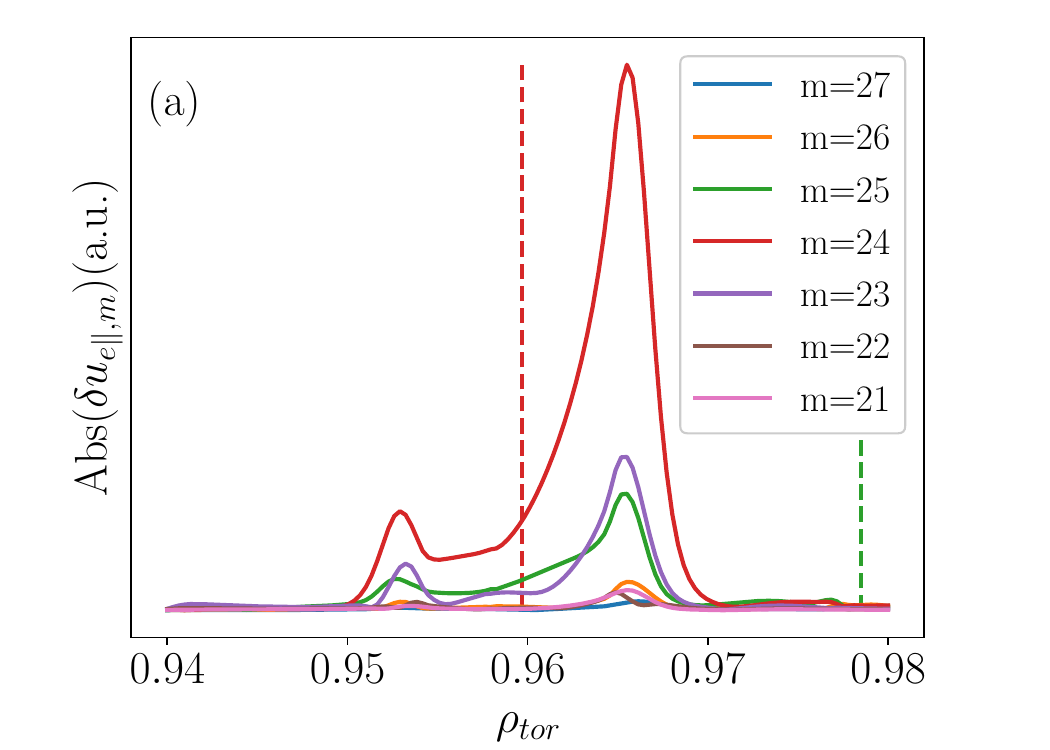}
    \includegraphics[width=0.32\textwidth]{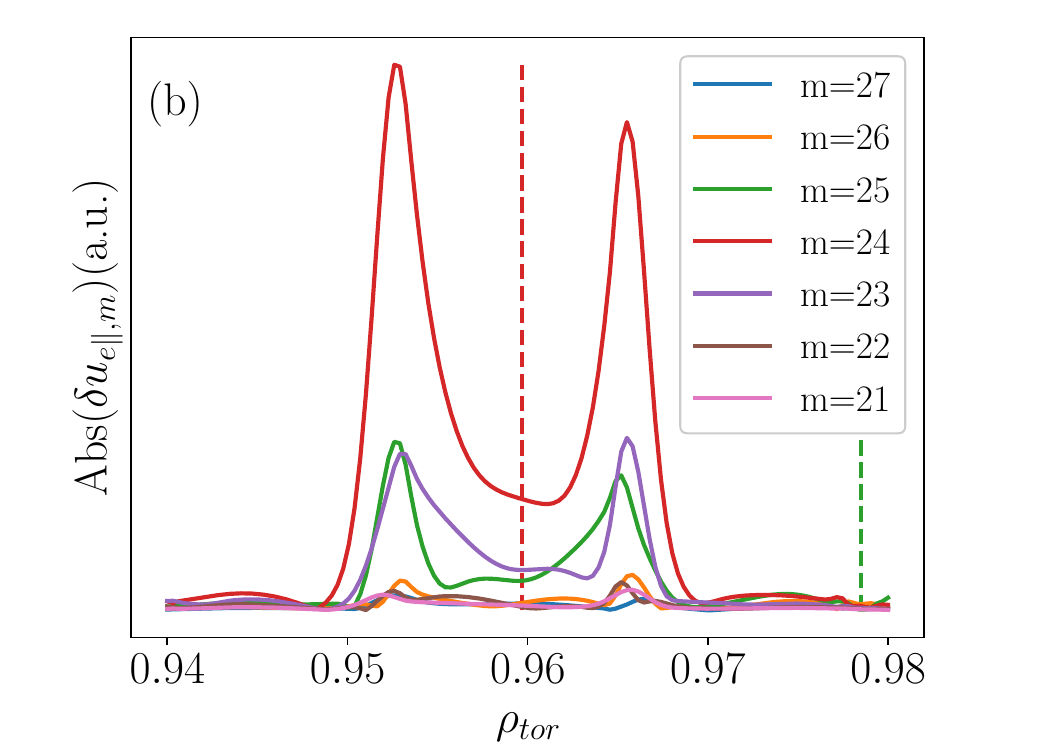}
    \includegraphics[width=0.315\textwidth]{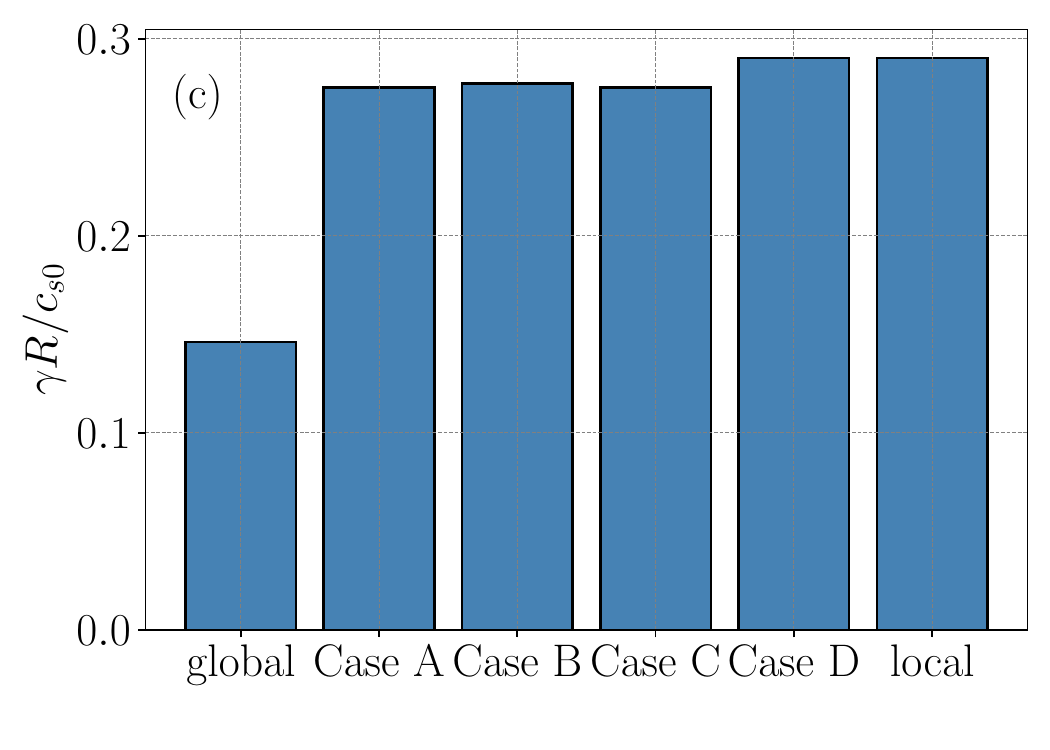}
  \end{center}
  \caption{(a) The current layer structure for $ n=6 $ global simulation. (b) The current layer structure for Case A. (c) The growth rates of $ n=6 $ simulations under different local assumptions. Case definitions: Case A uses local gradients profiles. Case B uses local gradients and metric coefficients profiles. Case C is the same as Case B but with $ L_x=\Delta_m $ and periodic B.C.. Case D incorporates all local assumptions.}
  \label{fig:mimic_n6}
\end{figure}

With the parameters in this study, $ \Delta _c$ in the pedestal is found to be $ \simeq 10\rho_{s0} $, which is comparable to the length scale of equilibrium profiles.
Consequently, both the growth rate and the structure of the current layer are significantly altered by the strong plasma non-uniformity, consistent with previous studies\cite{Chen2018, Larakers2021}.
For example, Fig. \ref{fig:mimic_n6}(a) illustrates the current layer structure from the $ n=6 $ global simulation in Fig. \ref{fig:ped_freq}.
When the profiles of $ n, T_e, T_i $, and their gradients $ R/L_A $ are fixed at their values at $ \rho _{tor}=0.96 $, the current layer structure exhibits significant modifications as shown in Fig. \ref{fig:mimic_n6}(b).
We refer to these profiles as `local gradients', as they correspond to the gradient profile assumption adopted in local simulations.
To comprehensively evaluate the influence of local approximations, the impacts of additional local approximations, including `local $ q $', `local metric coefficients', $ L_x=\Delta_m $ and periodic B.C., on the growth rate are also investigated and presented in Fig. \ref{fig:mimic_n6}(c).
Here, Case D incorporates all local assumptions in the global simulation and thereby mimics a local simulation.
From Case A to Case D, local assumptions are progressively imposed.
Through comparisons among these cases, the influence of each local assumption is quantitatively elucidated.
For low-$n$ modes, only the profiles of $ n, T_e, T_i $, and their gradients $ R/L_A $ exhibit a significant impact on the mode.

\begin{figure}[htbp]
  \begin{center}
    \includegraphics[width=0.4\textwidth]{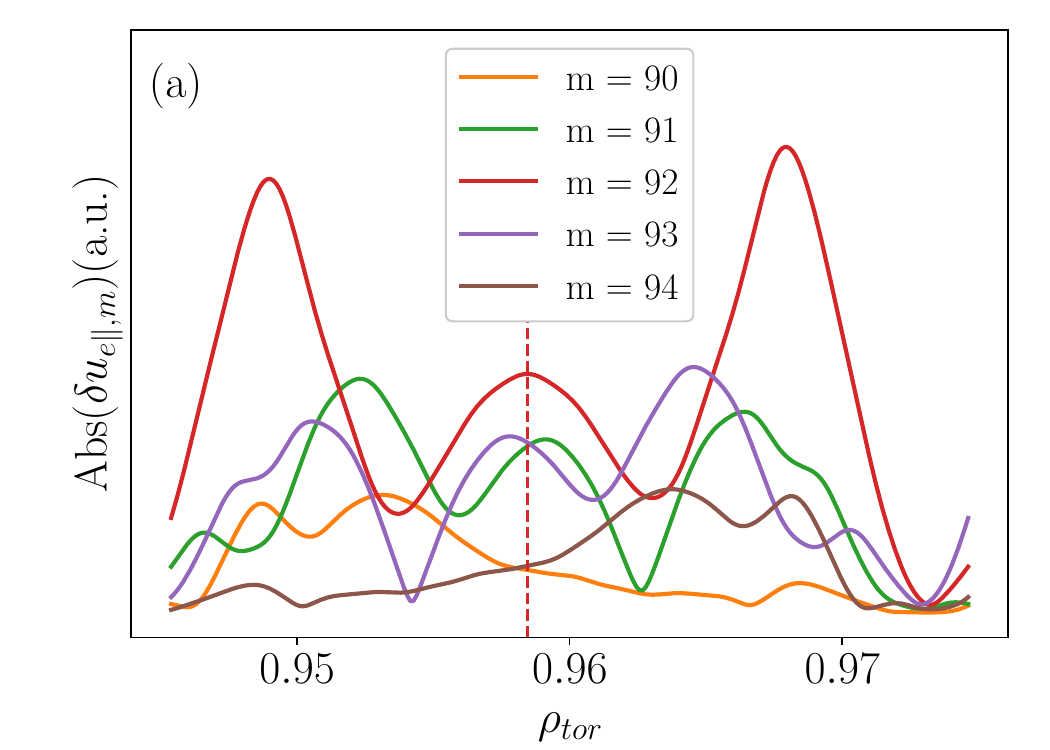}
    \includegraphics[width=0.4\textwidth]{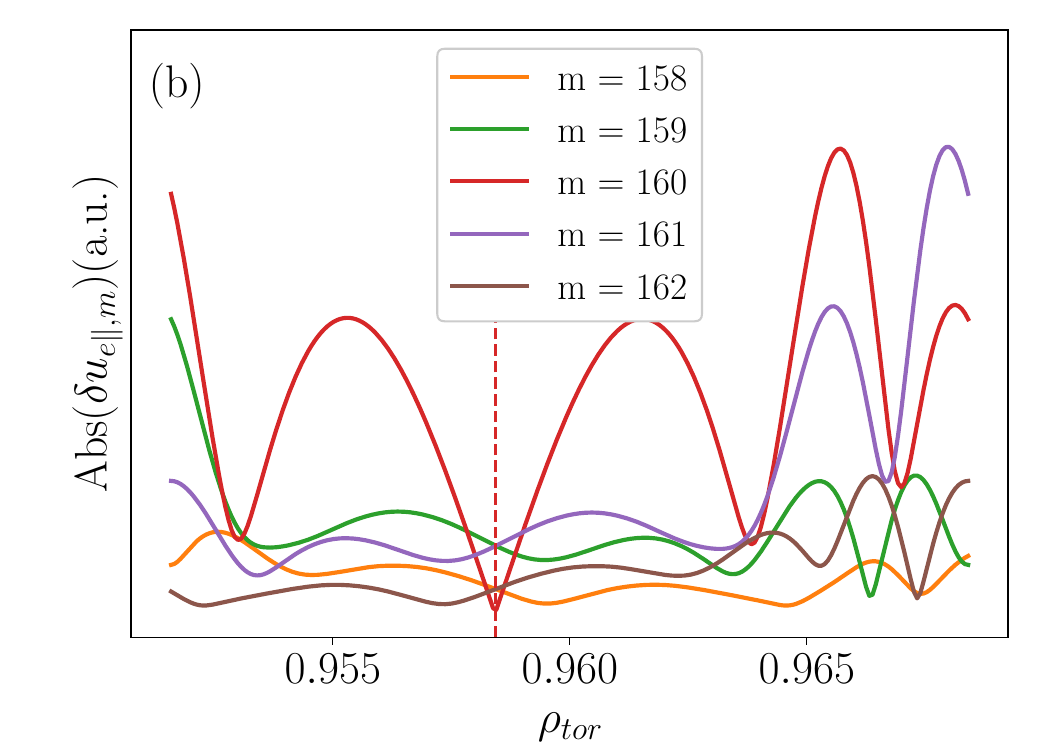}
  \end{center}
  \caption{The structures of current layers of (a) $ n=23 $ and (b) $ n=40 $ modes in local simulations.}
  \label{fig:ped_uapr}
\end{figure}

For high-$n$ modes, the overlapping between current layers from adjacent MRSs can occur.
Notably, $\Delta_c/\rho_{s0}$ in Eq. \eqref{eq:cl2} is independent of $n$, but $\Delta_m = 1/(\hat{s}k_y)$ decreases with $ n $.
When $\Delta_c > \Delta_m$, the resulting overlapping substantially change the mode.
% For high-$ n $ modes, it is worth noting that $ \Delta _c/\rho_{s0} $ in equation \ref{eq:cl2} is independent of $ n $, but $ \Delta _m = 1/\hat{s}k_y $, which depends on $ n $, can be narrower than $ \Delta _c $, which introduces the overlapping and mode coupling of current layers from adjacent MRSs, and qualitatively changes the mode.
For the typical pedestal parameters, the condition $ \Delta _c > \Delta _m $ yields
\begin{equation}
  k_y \rho_{s0} > \frac{1}{q} \sqrt{\frac{m_H}{m_e}} \Big/\left( \frac{R}{L_{ne}}+\frac{R}{L_{Te}} \right)\sim \mathcal{O}(0.1).
  \label{eq:crit}
\end{equation}
In the local simulations conducted here, the $ \Delta _c > \Delta_m $ regime is observed when $k_y\rho_{s0} > 0.2$ (i.e., $n > 30$).
Fig. \ref{fig:ped_uapr} compares the current layer structures for $\Delta _c < \Delta_m$ and $\Delta _c > \Delta_m$, where the latter case is characterized by current layers extending from adjacent MRSs into the reference MRS.
This extension facilitates toroidal coupling of current layers and substantially modifies the mode structure.
Accordingly, a significant rise in growth rates is observed in this regime (see Fig. \ref{fig:ped_freq}).
The coupling of current layers indicates that the MTM transitions into a toroidal mode, and the influence of the plasma non-uniformity is expected to be more significant.

\begin{figure}[htbp]
  \begin{center}
    \includegraphics[width=0.4\textwidth]{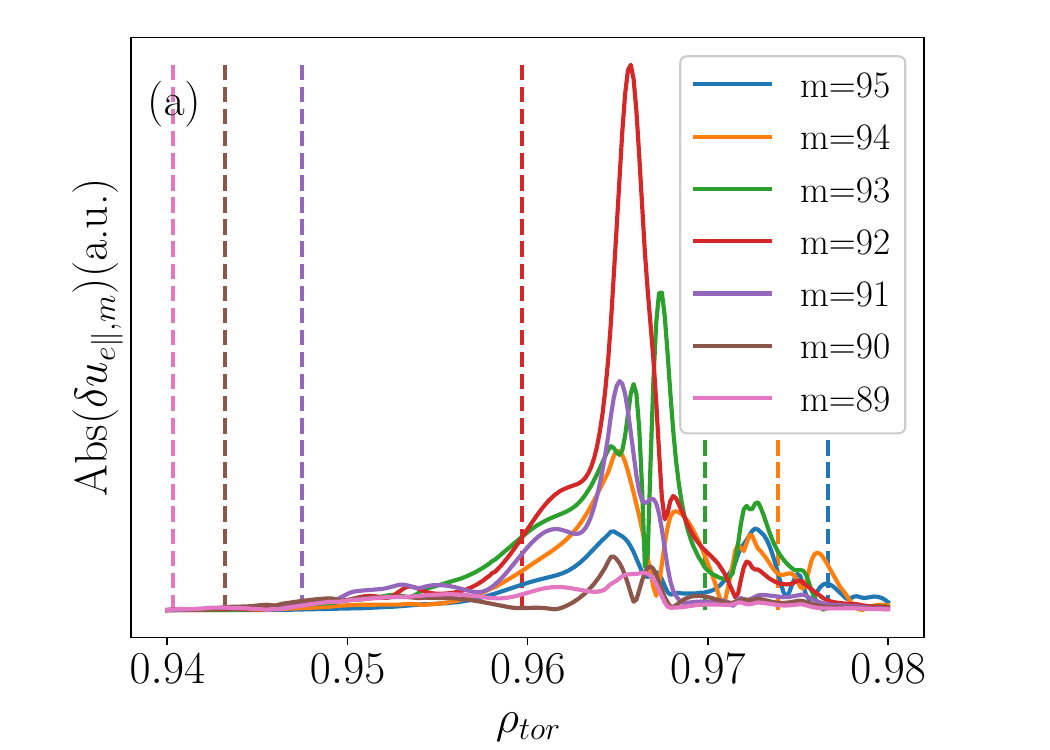}
    \includegraphics[width=0.4\textwidth]{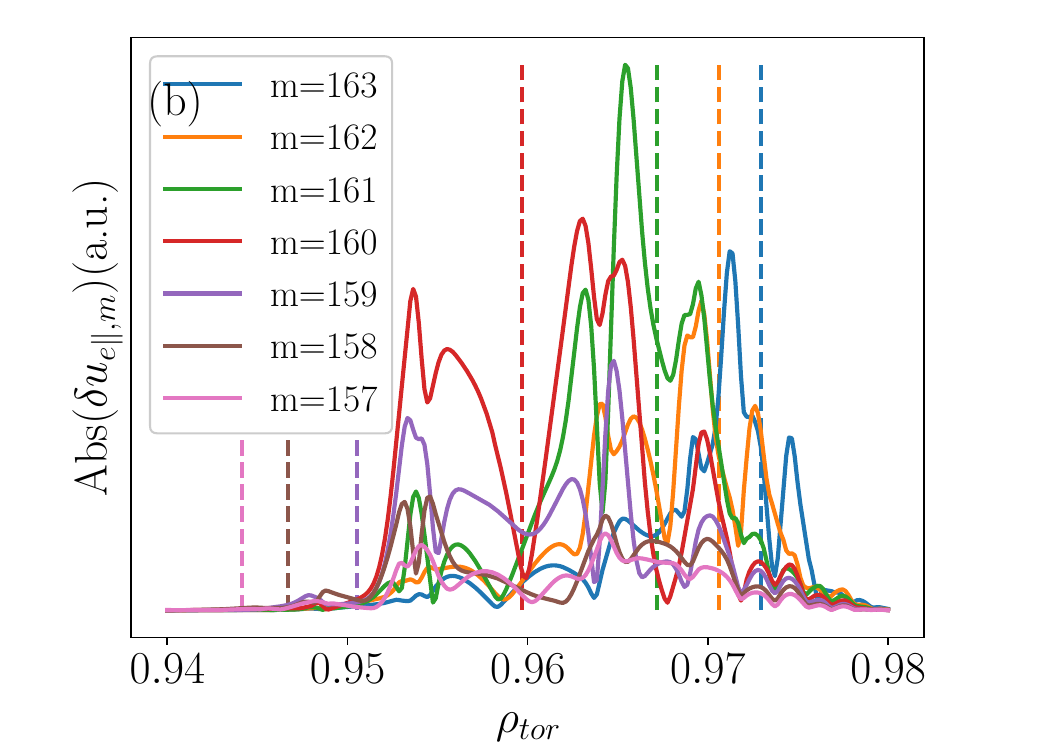}
  \end{center}
  \caption{The structures of current layers of (a) $ n=23 $  and (b) $ n=40 $ modes in global simulations. }
  \label{fig:ped_upar_g}
\end{figure}

\begin{figure}[htbp]
  \begin{center}
    \includegraphics[width=0.23\textwidth]{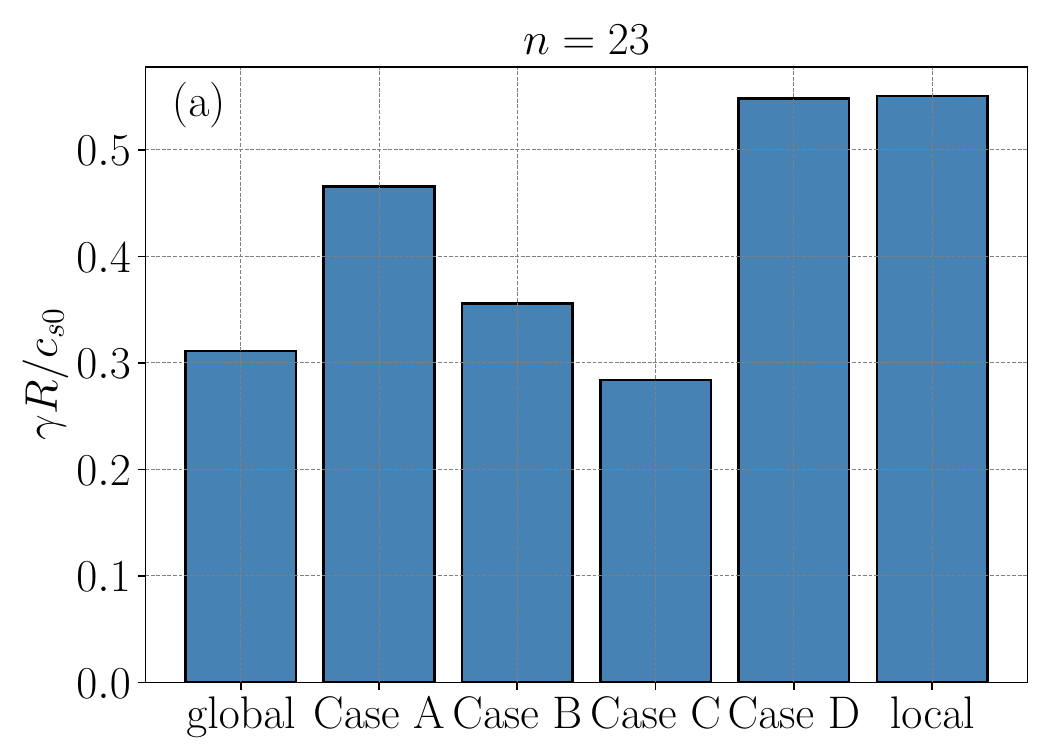}
    \includegraphics[width=0.23\textwidth]{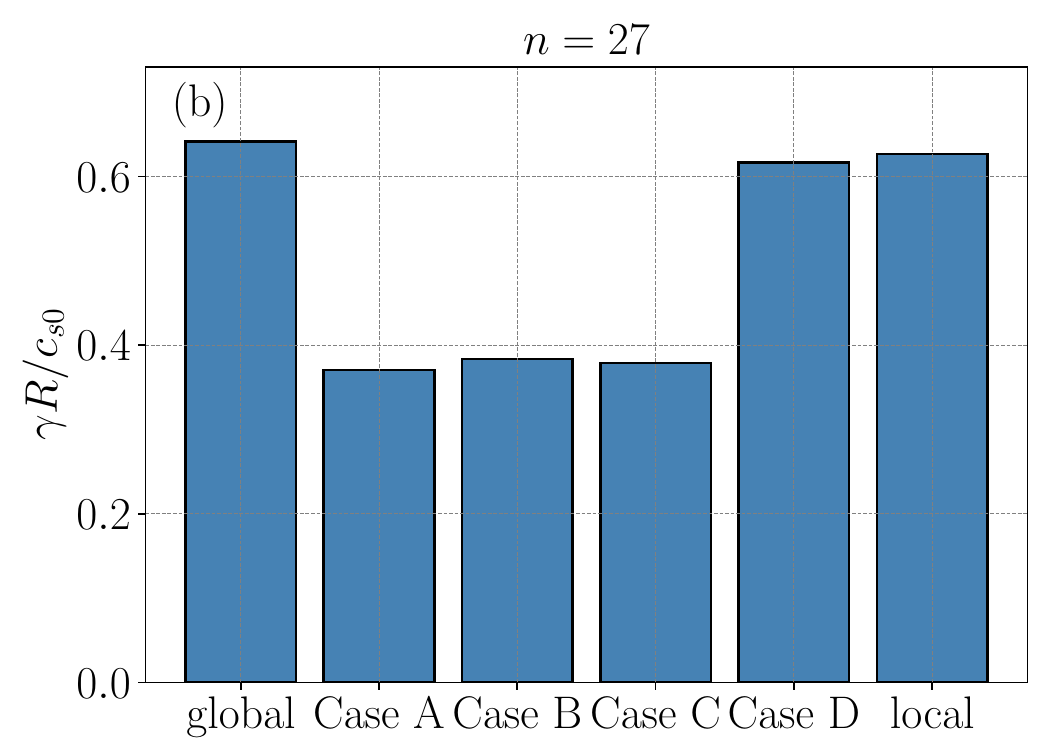}
    \includegraphics[width=0.23\textwidth]{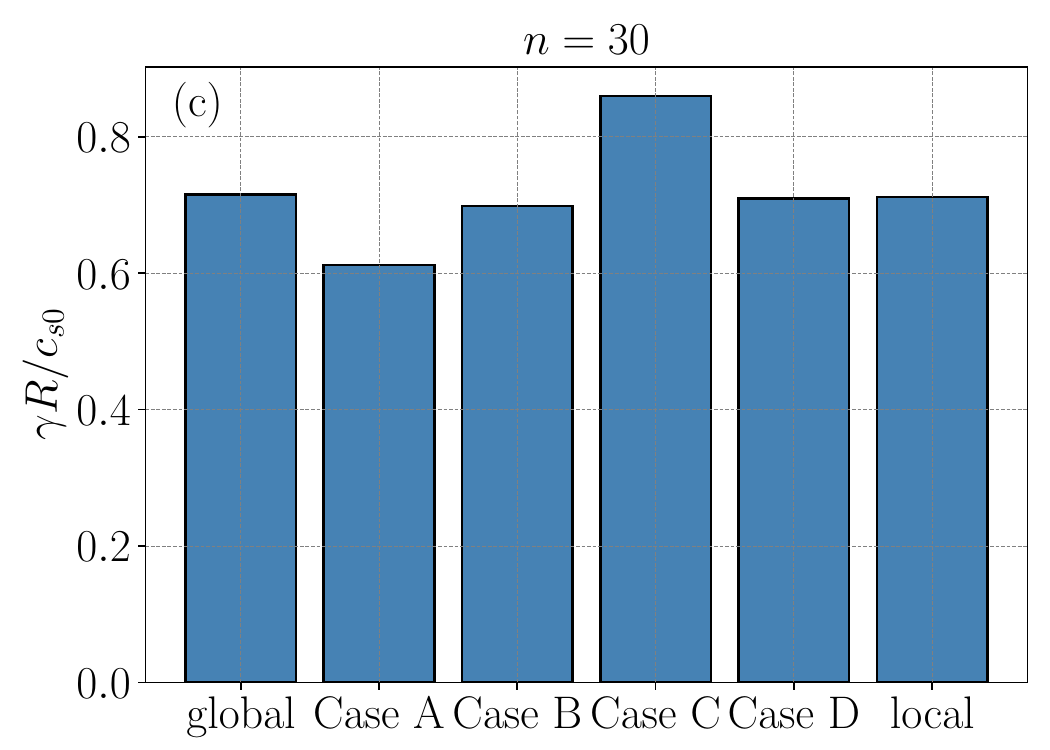}
    \includegraphics[width=0.23\textwidth]{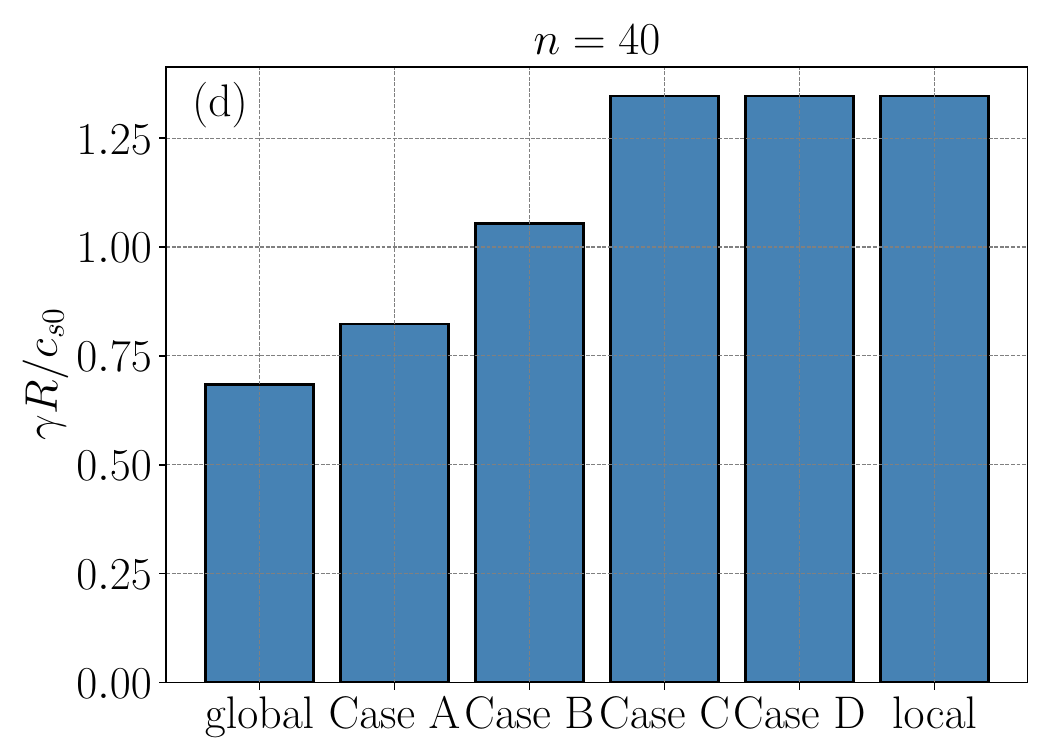}
  \end{center}
  \caption{The influence of local assumptions on high-$ n $ modes. The setup is identical to that in Fig. \ref{fig:mimic_n6}(c), except for $ n $.}
  \label{fig:mimic_high_n}
\end{figure}

In high-$ n $ global simulations, all unstable modes exhibit overlapping current layers.
For instance, Fig. \ref{fig:ped_upar_g}(a) shows the current layers for $n=23$, which couple with those in adjacent MRSs.
With increasing $n$ and narrowing $\Delta_m$, the toroidal coupling is enhanced,
leading to an expansion of the current layer envelope,
as demonstrated by the $n=40$ case in Fig. \ref{fig:ped_upar_g}(b).
These modes with coupled structures are expected to be further altered by the strong plasma non-uniformity,
as exampled in Fig. \ref{fig:mimic_high_n}.
In this regime, each local assumption significantly influences the mode properties without a clear trend as $n$ varies.
Consequently, the overlapping and coupling of current layers introduce a substantial distinction between global and local results for high-$n$ MTMs.
The apparent convergence or divergence of growth rate in Fig. \ref{fig:ped_freq} may be incidental.

\section{Summary}
\label{sec:Summary}
In this study, we use the GENE code to compare the local and global linear gyrokinetic simulation results of MTMs in tokamak plasmas.

The global simulations reveal a type of MTM characterized by the parity mixing structure, which is significantly destabilized by trapped electrons.
Unlike zero-ballooning-angle ($ \theta _k=0 $) MTMs, where the trapped electron response is weakened by bounce dynamics over an odd parity mode structure, this parity mixing structure corresponds to local simulations with a finite $\theta_k$ and avoids such cancellation.
In this regime, the poloidal harmonics are coupled, forming a broader MTM structure that spans multiple rational surfaces and renders the growth rate sensitive to profile variations. This finding underscores the necessity of finite $\theta_k$ effects in local MTM stability analyses.

It is found that the width of the current layer of MTMs, $ \Delta_c $, determines the importance of global effects, leading to global characteristics that are distinct from those of electrostatic drift wave instabilities\cite{Chen2018}.
In the core region, MTMs typically demonstrate highly local and global consistency and exhibit a slab-like feature characterized by weak coupling between harmonics.
This feature makes them insensitive to global profile variations in the core region, as the slab-like MTM stability is primarily determined by the equilibrium profiles within the $\Delta_c$ scale. 
However, in the pedestal region, $\Delta_c$ is broadened by the steep pressure gradient, quantitative deviations between local and global results emerge for low-$n$ MTMs when $ \Delta _c $ is comparable to the plasma pressure gradient scale length.
Moreover, for high-$n$ MTMs, $ \Delta _c $ can exceed the separation of MRS, $\Delta_m$.
This facilitates the overlapping of current layers from adjacent MRSs and their toroidal coupling, resulting into a radially extended mode structure.
This coupling mechanism accounts for the substantial discrepancies observed between local and global simulations.

\begin{acknowledgments}
The authors would like to thank Dr. Yang Chen for stimulating discussions.
This work was supported by the National MCF Energy R\&D Program under Grant No. 2024YFE03230300,
the National Natural Science Foundation of China under Grant Nos. 12375213,
the Natural Science Foundation of Sichuan Province No. 2025ZNSFSC0061,
the China National Nuclear Corporation ‘Young Talents’ Project No. 2024-QNYC-02,
the Innovation Program of Southwestern Institute of Physics (202301XWCX001),
the National MCF Energy R\&D Program under Grant No. 2024YFE03190004 and
the National Natural Science Foundation of China under Grant Nos. 12475215.
The simulations were performed on HPC Platform of Southwestern Institute of Physics and Tianhe new generation supercomputer of National Supercomputer Center in Tianjin.
\end{acknowledgments}

\section*{Data Availability Statement}
The simulation parameters and data that support the findings of this study are available from the corresponding author upon reasonable request.

\appendix
\section{Parity-based mode filtering in GENE}
\label{app:parity_filter}
In linear initial-value solvers such as GENE, the system typically converges to the most unstable eigenmode during the time evolution.
However, in specific parameter regimes, multiple microinstabilities can be unstable simultaneously, often obscuring subdominant modes.
To isolate and study these coexisting instabilities, a numerical parity filter module has been developed and integrated into the GENE code.
In the flux-tube simulation with the ballooning angle $\theta_k = 0$, the spatial parity of the electromagnetic fields along the equilibrium magnetic field line (characterized by the parallel coordinate $z$) is conserved as an exact symmetry. The modes can thereby be rigorously classified into two distinct parities:
\begin{itemize}
    \item \textbf{Ballooning parity} (typical of TEM and ITG modes): $\phi$ is symmetric (even) with respect to the outboard midplane ($z=0$), while the parallel vector potential $A_\parallel$ is antisymmetric (odd). Mathematically, this corresponds to $\phi(z) = \phi(-z)$, $A_\parallel(z) = -A_\parallel(-z)$.
    \item \textbf{Tearing parity} (typical of MTMs): The symmetry relations are reversed. That is, $\phi(z) = -\phi(-z)$, $A_\parallel(z) = A_\parallel(-z)$.
\end{itemize}

To numerically filter out the undesired parity, we impose the corresponding symmetry constraint at each simulation time step.
Specifically, the electromagnetic fields are first mapped to their mirror coordinates across the outboard midplane (i.e., $z \rightarrow -z$).
Because GENE's parallel domain decomposition typically distributes $f(z)$ and $f(-z)$ across different MPI processes, this operation requires a point-to-point MPI communication.
Pairs of MPI processes located symmetrically across the outboard midplane exchange their local spatial subdomains of the electromagnetic fields.
% that exchanges the local spatial subdomain of the electromagnetic fields between pairs of MPI processes located symmetrically across the outboard midplane ($z=0$). 
By incorporating the phase factor associated with the $k_x$ mode connections in the flux-tube boundary conditions, the reflected field $f(-z)$ is accurately reconstructed.

Depending on the targeted mode parity, the fields are then explicitly symmetrized or antisymmetrized. For instance, when the ballooning parity (TEM) filter is activated, the fields are updated
after the field equations are solved, according to:
% prior to the calculation of the right-hand side of the Vlasov equation according to:
\begin{align}
    \phi(z) &\leftarrow \frac{1}{2} \left[ \phi(z) + \phi(-z) \right], \\
    A_\parallel(z) &\leftarrow \frac{1}{2} \left[ A_\parallel(z) - A_\parallel(-z) \right].
\end{align}
Conversely, when the tearing parity (MTM) filter is activated, the signs of the reflected terms are inverted to enforce the opposite parity.

\begin{table}[htbp]
    \centering
    \begin{tabular}{l c c c c c}
        \hline\hline
        Filter Status & \texttt{mode\_parity\_ctrl} & \texttt{init\_cond} & $\gamma$ & $\omega$ & mode \\
        \hline
        Unfiltered & — & \texttt{almmt} & 0.079 & -1.200 & MTM \\
        Unfiltered & — & \texttt{alm}   & 0.085 & -0.880 & TEM\\
        Filtered   & \texttt{ballooning} & \texttt{almmt} & 0.085 & -0.881 & TEM\\
        Filtered   & \texttt{ballooning} & \texttt{alm}   & 0.085 & -0.881 & TEM\\
        Filtered   & \texttt{tearing}  & \texttt{almmt} & 0.079 & -1.200 & MTM\\
        Filtered   & \texttt{tearing}  & \texttt{alm}   & 0.079 & -1.201 & MTM\\
        \hline\hline
    \end{tabular}
    \caption{Validation of the parity filter at $k_y \rho_{s0} = 0.168$. The filter successfully and robustly isolates the tearing parity (MTM) and ballooning parity (TEM) modes independently of the selected initial condition (\texttt{init\_cond}). The growth rate $\gamma$ and frequency $\omega$ are normalized to $c_{s0}/R$.}
    \label{tab:parity_test}
\end{table}

To validate the effectiveness of this module in distinguishing different parities without altering their intrinsic physical properties,
% To validate the effectiveness and robustness of this parity filter,
representative test simulations were performed with the case in Fig. \ref{fig:stronger_gradient} (a) at $k_y \rho_{s0} = 0.168$, a parameter regime where the MTM and TEM possess closely competing growth rates. As demonstrated in table \ref{tab:parity_test},
without the parity filter, the apparent converged eigenmode is dependent on the initial condition:
% without the parity filter, the final converged eigenmode is highly sensitive to the initial condition:
an initial condition constructed with tearing parity (\texttt{almmt}) converges to an MTM, whereas an initial condition with ballooning parity (\texttt{alm}) converges to a TEM.
% The code converges to an MTM under the \texttt{almmt} initialization, but switches to a TEM under the \texttt{alm} initialization.
However, when the parity filter is applied, the solver successfully locks onto the intended physical mode, yielding identical growth rates and frequencies for a specific parity regardless of the initial conditions used.
% This validation proves that the diagnostic capability guarantees the unambiguous separation of the TEM and MTM branches throughout our analysis, thereby precluding numerical artifacts introduced by mixed parities.

\bibliography{ref}
\end{document}